\documentclass[prb,twocolumn,showpacs,preprintnumbers,amsmath,amssymb]{revtex4}
\usepackage{amscd,graphicx,color,colordvi,bm}

\newcommand{\nod}{\noindent}
\usepackage{verbatim}
\usepackage{epsfig}

\newcommand{\code}[1]{{\tt #1}}

\usepackage{rotating}

\begin{document}

\title{Digital Material: A flexible atomistic simulation code}

\author{Nicholas P. Bailey\footnote{Present address: CAMP, Building 307, Department of Physics, Technical University 
of Denmark, 2800 Lyngby, Denmark}}
\email{nbailey@fysik.dtu.dk}
\author{Thierry Cretegny}
\affiliation{LASSP, Department of Physics, Clark Hall, Cornell University, 
Ithaca, NY 14853, USA}

\author{Andrew J. Dolgert}
\author{Christopher R Myers}
\affiliation{Cornell Theory Center, Rhodes Hall, Cornell University, Ithaca, 
NY 14853, USA}

\author{Jakob Schi{\o}tz}
\author{Jens J{\o}rgen Mortensen}
\affiliation{CAMP, Building 307, Department of Physics, Technical University 
of Denmark, 2800 Lyngby, Denmark}

\author{James P. Sethna}
\author{Valerie R. Coffman}
\affiliation{LASSP, Department of Physics, Clark Hall, Cornell University, 
Ithaca, NY 14853, USA}

\date{\today}

\begin{abstract}
The complexities of today's materials simulations demand computer codes which
are both powerful and highly flexible. A researcher should be able to readily 
choose different geometries, different materials and different algorithms 
without having to write low-level code and recompile each time. We describe a 
molecular dynamics (MD) code, called Digital Material, in which we 
have sought to maximize flexibility without sacrificing efficiency. Our 
approach starts from the software engineering concept of Design Patterns and 
involves dividing the work of an MD simulation into well-defined components. 
The bulk of this paper is taken up with a detailed description of the different
components, their interfaces and implementations and the reasoning behind 
these. The level of detail is not at the line-by-line level, but at such a 
level that a reader could implement a similar code sharing the same design
 principles.
\end{abstract}


\maketitle 


\section{\label{introductionDM}Introduction: The complexities of today's 
simulations, as driven by multiscale materials modeling}

Simulations of materials continue to become increasingly complex,
driven by the need for greater modeling fidelity and
the opportunities provided by advances in available computational
resources.  In recent years, this complexity seems to have advanced at
an even faster rate, as powerful-but-unwieldy parallel computing
platforms have become widely available for high-performance
computation, and as researchers have reached across disciplinary
boundaries to address the multiscale nature of material 
behavior\cite{Bulatov/others:1999}.

Material structures and phenomena are inherently multiscale, so the
desire for greater realism in materials modeling has driven a growing
interest in multiscale materials modeling techniques.  In some cases,
these techniques explicitly link together disparate numerical models
(at different length and/or time scales) to form hybrid meta-models.
In other cases, simulation results from one scale are implicitly
incorporated into computational models at other scales (e.g., in the
form of constitutive descriptions).  Investigation of material
behavior across scales can involve treatment of more complex
simulation geometries, boundary conditions, constitutive models, and
numerical algorithms.  Furthermore, optimal models and/or numerical
methods are in many cases not yet known, and need to be discovered
through numerical experimentation.  All of these trends conspire to
suggest a need for more sophisticated software frameworks to support
the generation of complex material models, the construction of
compound and hybrid numerical methods, the structuring of code for
high performance on modern supercomputers, and the flexible control
and interrogation of simulations and data.

Complexity in the multiscale investigation of materials can arise from
many sources, and has many implications for the software development
process.  Whereas much of atomistic modeling to date has involved
relatively simple simulation geometries, the desire to provide input
to processes active at larger scales (e.g., plasticity and fracture)
increasingly requires construction of atomistic models with more
complicated geometries, involving, say, sets of interacting
dislocations or a grain boundary of a specified misorientation.
Furthermore, extracting useful information from small-scale atomistic
simulations for use in larger scale theories or models requires
careful treatment of boundary effects.  This has led to the
development of hybrid numerical methods for finding the structure of
atomistic defects (e.g., dislocation cores); these hybrid methods
can involve both atomistic and continuum degrees of freedom which are
simultaneously acted upon.

Our efforts in developing software frameworks for materials modeling
fall under the general rubric of Digital Material (DM), which connotes 
both a general approach to software development for materials simulation
and specific software systems for particular types of 
simulation\cite{Myers/others:1999,Myers/others:2001}.
Our focus in this paper is on the atomistic modeling system that we have
developed\cite{DMCode}. Prior to describing
the specific details of the DM atomistic modeling system, we present some
of the high-level design and implementation goals that characterize the
DM effort broadly.

\subsection{Digital Material design goals}

We aim to build a system that is flexible, expressive, and extensible,
while not sacrificing computational performance.  We believe it is
important to support composition of many computational modules, both
to enable the construction of hybrid models and methods, and to
facilitate the development of simulation software.  Fortunately, there 
are a number of recent software engineering developments which we can
exploit to build such a system.

{\it Design patterns}\cite{Gamma/others:1995} represent an important set of
object-oriented design techniques to have emerged in the last decade.  These 
patterns address the collaborations among computational objects, in such a way
as to support software change and reuse.  An important element of these
design patterns is that they aim to support additive rather than invasive
change.  That is, if a new piece of functionality is desired, it is
preferable to be able to add (plug in) a new module rather than change 
(rip up) an existing one.  Developing the correct decomposition of 
desired functionality to facilitate change of this sort is one of the
central tasks in building such a system.  As such, describing such a 
decomposition lies at the heart of this paper.  In this introduction,
we introduce some of the more general patterns which guide the overall
structure of the system. It is important to note the tensions that arise 
between design patterns as
they typically are used and traditional high performance scientific
computations.  Design Patterns emphasize indirection and delegation, whereas
scientific computation typically avoids such techniques because of performance
 concerns. One of our goals is to define and develop
a new set of design patterns for scientific computing, which leverage
useful software design principles without unduly sacrificing computational
performance.

Materials simulations typically involve one or more material
``samples'' (instantiations of the relevant modeling
degrees-of-freedom, e.g., atoms, grains, displacement fields, etc.) 
which are acted on by some model of a physical processes (e.g.,
applying loads, following a time evolution).  We have therefore chosen
to separate our description of material samples from the ``movers''
that act to modify those samples.  This allows us to identify a
material state independently of the models used to modify that state,
and to switch different sorts of movers in and out as we develop
complex models and algorithms.  (In a similar fashion, computational
probes which interrogate the state of a material sample are also
separated out as ``observers'' of the underlying state.)  Furthermore,
we have chosen to subdivide the description of a material sample into
one or more sets of geometric entities with associated sets of
attributes. In particular, our \code{ListOfAtoms} is \textit{not} an array of
 objects
of a class \code{Atom}, but rather is an array of positions, plus an array of
velocities, etc. This is useful for several reasons.  First, for reasons
of performance, it is necessary to act on aggregates of data in tight
numerical kernels (the ``inner loops'') without the cost of
higher-level overhead and control.  In atomistic modeling, the
positions of the atoms (which constitute the geometry of the sample)
are accessed to compute neighbor lists and forces, and we wish to be
able to access that geometric information independently of other
attributes (e.g., masses or velocities) for optimal performance.
Second, there are other operations (e.g., visualization, or
computation of local atomic coordination number) where only geometric
information is necessary, and we would like to be able to extract that
information without striding over all other atomic attribute data.
Finally, inherent in many approaches to multiscale modeling is the
need to treat different material objects in different contexts at
different scales: a dislocation line, for example, is a collective and
emergent feature in an atomistic simulation, while being explicitly
represented as part of the computational model in a dislocation
dynamics simulation.  The power of multiscale modeling often lies in
the ability to factor a complex description (e.g., a constitutive
model at one scale) into a geometric piece and a different set of less
complex descriptions: the quasicontinuum method\cite{Shenoy/others:1999}, for
 example, replaces
complex continuum constitutive descriptions of solids with an
alternate description, involving the mutual self-organization of
collections of atoms (a geometric structure) interacting via
interatomic potentials (a less complex constitutive description).

Another important software engineering development which has had a
significant impact on our research is the growing use of high-level
interpreted scripting languages to control and steer compiled
numerical simulation frameworks.  Prominent examples demonstrating the
value of this approach include SPaSM,\cite{SPaSM,Beazley/Lomdahl:1997} 
a system for
molecular dynamics simulations of solids, and the Molecular Modeling
Toolkit (MMTK)\cite{MMTK}, for biomolecular simulations. Bahn and 
Jacobsen\cite{Bahn/Jacobsen:2002} describe the use of such a language to 
interface with a legacy electronic structure code, and have dealt with many of
 the same issues we address here. Like these
other projects, we use the Python programming language to develop
high-level interfaces to our simulation kernels, and to glue together
applications composed from disparate pieces (for numerical algorithms,
data storage, visualization, graphical interfaces, etc.).  A
lightweight, programmable interface layer like Python supports our
need for flexible prototyping, control and interrogation, without
impacting low-level computational performance.

\section{Components}\label{components}


Others have described the various computational techniques and algorithms which
constitute the standard molecular dynamics 
``toolbox''\cite{Allen/Tildesley:1987}. This is not our focus. Rather, in this 
section, which makes up most of the paper, we systematically describe
the components we have identified as being logically separate pieces of a
molecular dynamics code.  We start with the main data structure, 
\code{List\-Of\-At\-oms}, and continue with the components that define the
geometry of the simulation.  The \code{AtomsInitializer} is responsible for
populating the \code{ListOfAtoms}, usually with a crystal structure of atoms
that fills a simple geometry such as a sphere or rectangular prism.  The
\code{BoundaryConditions} class is responsible for maintaining free or periodic boundary
conditions, where appropriate.  The \code{Transformer} class is often used to alter
the geometry of the atoms, to add a void or dislocation, for instance.  The next
set of components perform functions related to the dynamics of the simulation.
Because interatomic potentials generally have a finite cutoff distance, it is useful
to be able to find all of the atoms within the cutoff of a given atom.  This
function is performed by the \code{NeighborLocator} component of DigitalMaterial.
We have a \code{Potential} component, which is responsible for calculating the
  potential energy and forces for the atoms.  The \code{AtomsMover} then uses the
  \code{Potential} class to either perform molecular dynamics or energy
  minimization. The optional \code{Constraint} component can be used to restrict
  the motion or forces on a group of atoms.  Finally, we have an \code{Observer}
  component which is responsible for recording the various types of results
  produced by a simulation.

In section \ref{infrastructure} we describe aspects of the code which are
considered ``infrastructure'': serialization, parallelization and graphics.
The programming language we have used for most of the code is C++. Although we
 wish to emphasize that the philosophy underlying the development of this code
is independent of language, it will be useful to have a specific framework for
the discussion of classes, etc., and thus we will sometimes make reference to
C++ constructs. When referring to function names, etc., we will 
generally omit the complete type information that is actually required
 in the code, for clarity. The same for the tables of method names
provided for the abstract classes associated with the principal components.


\subsection{ListOfAtoms}\label{listofatoms}

\subsubsection{Responsibilities}

The primary responsibility of a \code{ListOfAtoms} is to store and provide 
access to the current state and properties of the atoms in the simulation. The
 base class stores simply the positions. The secondary responsibilities of a 
\code{ListOfAtoms} are notifications:

\begin{enumerate}
\item It ensures the validity of its current state by passing on
changes to the \code{Constraint}s (section~\ref{constraint}) and 
\code{BoundaryConditions} (which may for example project the atoms back into 
the supercell; see section~\ref{boundaryconditions}),
\item It warns the \code{NeighborLocator} (section~\ref{neighborlocator}) of 
changes in state (thus prompting a check e.g. of whether its neighbor list 
needs to be rebuilt),
\item It acts as a Subject which, when prompted by the user, notifies a
stored list of Observers (for visualization, or any kind of analytical
measurement, etc.; see section~\ref{listofatomsobserver}).
\end{enumerate}


\begin{table*}\label{ListOfAtomsMethods}
\begin{center}
\caption{Methods for \code{ListOfAtoms}.}
\begin{tabular}{|l|l|}
\hline
ListOfAtoms & (contd.)\\
\hline
GetNumber() & GetBoundaryConditions()  \\
GetShape() & SetBoundaryConditions()  \\
SetShape() &  GetNeighborLocator() \\
RemoveAtoms() & SetNeighborLocator()  \\
Copy() & SetConstraint()  \\
Merge() & Clear()  \\
Attach() & IsLeaf()  \\
RemoveBranch() & NumberOfBranches()  \\
double GetMass() & GetBranch()  \\
void SetMass() &  NumberOfLeaves() \\
GetCartesianPositions() & GetLeaf()  \\
SetCartesianPositions() & SetUpperNode()  \\
GetCartesianPosition() & AddObserver()  \\
SetCartesianPosition() & RemoveObserver()  \\
IncrementCartesianPosition() & Notify() \\
IncrementCartesianPositions() & AdjustPositions()  \\
GetCartesianVelocities() & AdjustForces()  \\
SetCartesianVelocities() & AdjustForceIncrements()  \\
GetCartesianVelocity() & AdjustVelocities()  \\
SetCartesianVelocity() & NumberGeneralizedCoordinates()  \\
IncrementCartesianVelocities() & SetGeneralizedCoordinates()  \\
IncrementCartesianVelocity() & GetGeneralizedCoordinates()  \\
IncrementCartesianMomenta() & IncrementGeneralizedCoordinate()  \\
GetKineticEnergy() & CalculateGeneralizedForces()  \\
ScaleCartesianVelocities() & DerivativesWRTGeneralizedCoordinates() \\
\hline
\end{tabular}
\end{center}
\end{table*}

\subsubsection{Examples}

Velocities are not always necessary in an atomistic simulation (e.g., for
energy minimization), but are often important so our main derived class, 
\code{Dynamic\-ListOfAtoms}, adds velocities
to the state. A more sophisticated example comes from our implementation of the
quasicontinuum method, which mixes MD with finite elements. Here we have 
derived slave and master atoms classes, so that each slave atom can determine
the element it belongs to and each master atom has its own weighting 
factor, e.g. for the energy calculation.

\begin{figure}[thbp]
\begin{center}
\epsfig{file=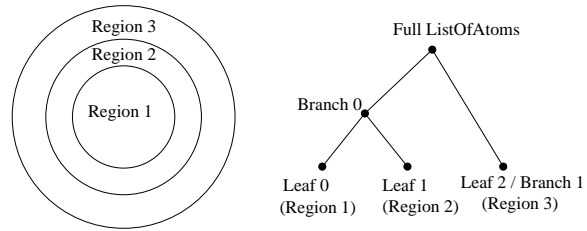, width = 3. in}
\caption{\label{branchesIllus}Illustration of the tree structure for a 
ListOfAtoms. Here the application requires the atoms to be grouped into three
``regions'' as shown, where Regions one and two at times will be treated 
together. This structure is reflected in the data structure shown on the 
right. }
\end{center}
\end{figure}

\subsubsection{Implementation, Efficiency, Flexibility}

Cache performance led us to favor arrays of native types for storage. The cache
on current processors loads an entire line of contiguous data whenever
a value is fetched, so storing all the position information in
separate array (rather than in an atom class, mingled with velocities
and other attributes) reduces the number of cache misses. 

Pipelining leads us to store all atoms of a given type together (and
atoms subject to the same constraint together). The concurrent
processing of multiple sequential instructions is easiest for the
compiler to optimize when simple tasks are repeated in regular
patterns (facilitating loop unrolling, ...). Control statements (like
``if (atom.type()==\ldots'') typically cause pipelines to stall. By
putting the atoms of a given type together, these control statements
are implemented once per type outside the loop over atoms.

We implement \code{ListOfAtoms} as a tree structure, with each atom type
(subject to each kind of constraint), on its own leaf(see 
Fig.~\ref{branchesIllus} for an example). By making each
branch and leaf of the \code{ListOfAtoms} itself a \code{ListOfAtoms}, 
other classes (e.g., for graphics and correlation functions)
can work on subsets of the atoms without modification. This is an example of
the Composite design pattern\cite{Gamma/others:1995}. An example is illustrated
in Fig.~\ref{branchesIllus}. \code{Constraint}s can 
be applied to sublists of
\code{ListOfAtoms} without the constraint class being aware of the surrounding
atoms. We also chose to store the data for the atoms in a tree 
structure (the \code{DMArray} class), which mirrors the tree structure of 
\code{ListOfAtoms}. 

\subsubsection{Alternative Choices}

Our tree-structure array class has ended up being quite complex. Some of
the complexity is needed because of the need to support parallel
processing. The entries in temporary force arrays in the \code{AtomsMover}s,
for example, need to be registered with the base class so that their
entries are automatically transferred to other processors when
the atoms cross processor boundaries. Some of the complexity,
however, could have been avoided by storing the data for all the
sublists contiguously in memory. This has the disadvantage that each
time atoms migrate the entire list of atoms must be shuffled up or
down to make room. On the other hand, the current implementation
demands extra overhead for computing the address of the neighboring
atoms in force loops.

The \code{DMArray} class also makes heavy use of templates. In an application
where all attributes of atoms were of type \code {double} (or
\code{double[DIMENSION]}) this could have been avoided, making the
class easier to read (but reducing the flexibility). There are also
many other freely available templated C++ array classes (such as
BLITZ++\cite{BLITZ++:1997}), but they do not support the kinds of hierarchies 
in storage that we needed here.

Finally, there are other models for notification that we could have
used. Instead of having \code{ListOfAtoms} call \code{BoundaryConditions} and
\code{Constraint}s and notify \code{NeighborLocator} these responsibilities 
could have been left to a ``MotherBoard'' simulation class, which would
function as something like the Mediator design pattern\cite{Gamma/others:1995}
to encapsulate communications among the various objects in the system, 
and thereby better insulate those objects from one another.

\subsection{AtomsInitializer}\label{atomsinitializer}


While not as crucial as \code{Potential} and \code{Neighbor\-Locator}, a 
certainly significant component of our software is the \code{AtomsInitializer}.
 Since for the most part,
initialization is something that occurs once in a simulation, efficiency is not
the key issue here. The purpose of \code{AtomsInitializer} classes is to save 
{\it user time} rather than {\it computer time}. If one were to have to think
 about the coding details of getting
orientations of crystals and axes right every time one wanted a new simulation,
one might be tempted not to change the simulation geometry very frequently. The
key benefit of having a set of Initializer classes implemented is that with
 very
little work---generally a few lines of a python script---one can set-up a wide
variety of configurations. The flexibility comes from separating the lattice to
be used for filling from the shape defining which region of space is to be
filled. Further flexibility derives from the facility to compose different
shapes in various ways.

\begin{figure*}[thb]
\begin{center}
\epsfig{file=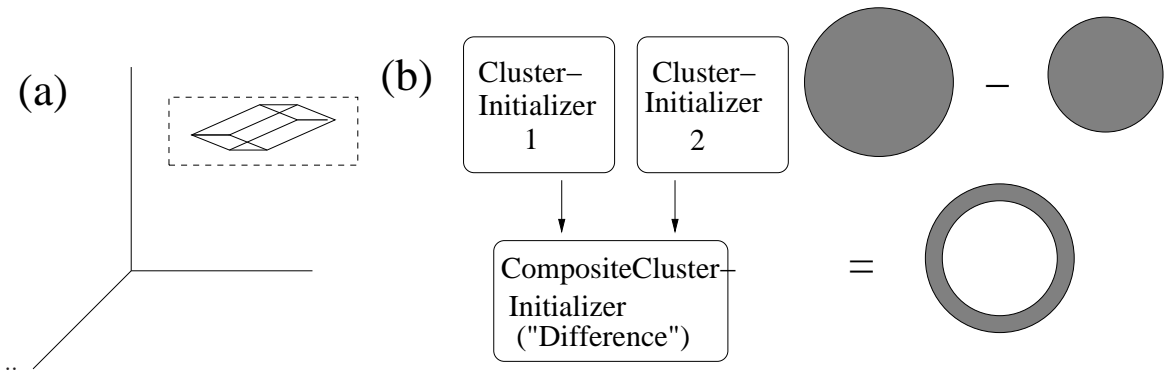, width = 6 in.}
\end{center}
\caption{\label{InitializerIllus} Examples of \code{ClusterInitializer}s (a)
 Filling a parallelepiped with atoms. The dotted line indicates the volume of
 space to be searched for candidate positions. This must include all of the 
intended region and ideally as little as possible extra. (b) Using
\code{CompositeClusterInitializer} to create an annular region as the 
difference of two circular regions.}
\end{figure*}

\begin{table}\label{AtomsInitializerMethods}
\begin{center}
\caption{Methods for \code{AtomsInitializer}.}
\begin{tabular}{|l|l|}
\hline
AtomsInitializer & ClusterInitializer \\
\hline
Create() & SetCenter() \\
GetCenter()) & GetBravaisLattice() \\
GetMaxSize() & SetBravaisLattice() \\
\hline
\end{tabular}
\end{center}
\end{table}

\subsubsection{Responsibilities} 

An \code{AtomsInitializer} must possess a \code{Create()} function. An empty 
\code{List\-Of\-At\-oms}
is passed to the Create function after which it is no longer empty but has a
number of atoms and a set of positions and velocities determined by what type
of initializer it is, and what parameter values were passed to it. Typically 
one wants to fill regions of space with atoms. For solids typically studied
 using atomistic modeling these are in a
crystalline array. This necessitates a lattice class of some sort. Our lattice
class is called \code{Brav\-ais\-Latt\-ice\-With\-Bas\-is} and gives a general 
crystalline
lattice, with
arbitrary lattice vectors and arbitrary number and positions of atoms within a 
unit cell. Subclasses with specific lattice vectors and bases have been
defined for the common lattices: \code{SimpleCubic}, \code{FCCLattice}, 
\code{DiamondLattice}
etc. Methods include rotations and translations, operations to convert between
real coordinates and lattice coordinates etc.

The most important type of Initializer is the \code{Cluster\-Initializer}. This 
is a base class for several different concrete subclasses (see below, and 
Fig.~\ref{InitializerIllus}, for examples). The base class provides the
 \code{Create()} function, and takes a \code{BravaisLattice} as a
constructor argument. The subclass must provide an \code{Inside()} function, 
which takes a point in space and returns a boolean value if the point is inside
 the region to be filled.

If the initial state of a simulation is not a homogeneous crystal occupying 
some region of space, but perhaps has a strain field of some sort applied, or
 has one 
or more defects (dislocations, notches, cracks, vacancies, etc.) one uses an
initializer such as a \code{ClusterInitializer} and subsequently applies a 
\code{Transformer}
to the \code{ListOfAtoms}. \code{Transformer}s are designed to make geometrical
 and topological changes to an already existing \code{ListOfAtoms}; see section
\ref{transformer} for more on \code{Transformer}s.

\subsubsection{Implementation, Efficiency, Flexibility} 

Since, as we have said, initialization is only performed once, efficiency is
not as crucial as with the \code{NeighborLocator} or the \code{Potential}. 
However there is 
one important place where some thought can be usefully spent in order to reduce
start-up time. The \code{Inside()} of a \code{ClusterInitializer} can only say
 whether a 
given position is or is not within the region to be filled; it does not by
itself give suggestions for candidate positions. Since we cannot loop through
every lattice position in space, we need a reasonably good estimate of a 
bounding region which has a simple shape, which can be looped over, passing 
each lattice position to the \code{Inside()} function to be tested. Thus each
 subclass must implement a function, called \code{SetMaxSize()}, which 
determines an appropriate
range of lattice vectors to be looped over. Care has to be taken to ensure that
 the region of space thus defined definitely includes the region to be filled, 
regardless of how skewed the lattice vectors are. Another point of care here is
 the proper
treatment of positions on the boundary itself. This is important for example
when filling a rectangular region with atoms and applying periodic boundary
conditions to that region---it is very easy to end up having two different
atoms occupying sites which are equivalent by periodicity, so the two atoms
are effectively on top of each other (this is bad!). We treat these situations
by not including sites near the ``negative'' boundary, while including sites
near the positive one. This becomes tricky if the boundary faces correspond to
high-index (low symmetry) crystal planes, and then it can be simpler to err 
towards double-placing atoms and subsequently removing any extra atoms.

Further flexibility is achieved with the 
\code{Com\-pos\-ite\-Clus\-ter\-In\-it\-ial\-iz\-er} 
class. This allows one to have as the region to be filled the union, 
intersection or difference of two other regions, defined by other 
\code{Clus\-ter\-In\-it\-ial\-izer}s. This is
very useful in mixed atomistic-continuum simulations where the central region
of atoms is a sphere or disk or cylinder, and this is to be surrounded by a
shell or annulus of constrained atoms. The \code{Inside()} function of the
\code{Com\-pos\-ite\-Clus\-ter\-In\-it\-ial\-iz\-er} performs the appropriate 
boolean operation on
 the results from the \code{Inside()} functions of the other 
\code{Clus\-ter\-In\-it\-ial\-izer}s.

\subsubsection{Examples} 

Examples include

\begin{enumerate}
\item \code{SphericalClusterInitializer}: fill a sphere of given center and 
radius.
\item \code{RectangularClusterInitializer}: fill a rectangular region with 
sides parallel to coordinate axes given center and lengths of the sides.
\item \code{CylindricalClusterInitializer}: fill a cylinder of given center, 
radius and height.
\item \code{PolyClusterInitializer}: fills a simple (by default convex) polygon
 or polyhedron in two or three dimensions respectively.
\item \code{ParallelepipedClusterInitalizer}: fills a parallelepiped of 
specified edge vectors and center.
\item \code{AtomicSurfaceInitializer}: this was designed for a more specific
application, measuring surface energies. It sets up layers of atoms of 
specified thickness and orientation relative to crystal axes. It also sets the
 lengths of the boundary conditions appropriately.
\end{enumerate}

\subsection{BoundaryConditions}\label{boundaryconditions}


\begin{figure}[thb]
\begin{center}
\epsfig{file=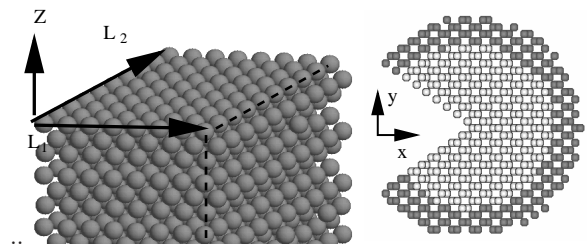, width = 3 in.}
\end{center}
\caption{\label{BCIllus}Two examples of boundary conditions. The system on the
left is a 111 surface of Cu. It uses \code{SkewPeriodic\-BoundaryConditions} with
vectors $\vec L_1$ and $\vec L_2$ defining the in-plane periodicity. The third
vector is perpendicular to them but periodicity is turned off so that it does
not play a role. In the system on the right, representing a notched single
crystal of Si, the opposite situation occurs: periodicity is applied only in the
out of plane ($z$)direction, so that in the plane there are technically free
boundary conditions. However, the outer layer of atoms (dark gray) is
constrained and acts as the effective boundary for the inner atoms.}
\end{figure}

\begin{table*}\label{BoundaryConditionsMethods}
\begin{center}
\caption{Methods for \code{BoundaryConditions}.}
\begin{tabular}{|l|l|}
\hline
BoundaryConditions & PeriodicBoundaryConditions \\
\hline
EnforceBoundaryConditions() & SetLength() \\
DifferenceBoundaryConditions() & GetLength() \\
\hline
\end{tabular}
\end{center}
\end{table*}

\subsubsection{Responsibilities} 

The boundary conditions have basically two clear responsibilities.
\begin{enumerate}
\item They must make sure that the position of every atom satisfy the boundary
conditions: we later refer to this task as \code{EnforceBoundaryConditions()}.
\item They must determine the actual separation between a pair of atoms,
identifying their closest images through the boundary conditions.  We
call this \code{DifferenceBoundaryConditions()}.
\end{enumerate}

\subsubsection{Examples} 

The traditional type of boundary conditions employed in an MD simulation is
\code{PeriodicBoundaryConditions}. We have implemented a few variants of these,
offering for example the choice to switch off wrapping in certain directions,
or allowing a general parallelepiped shape rather than a rectangle---this being
useful for studying systems under shear for example, or studying surfaces of
various orientations. Two examples are illustrated in Fig.~\ref{BCIllus}.

The simplest kind of boundary conditions are the kind that do nothing at all,
known as \code{FreeBoundary\-Conditions}. This is useful in simulations where 
a region of freely moving atoms is surrounded by a region of fixed or 
constrained atoms which provide effective boundaries, or for simulations of 
clusters.

\subsubsection{Implementation, Efficiency, Flexibility}  

Because of the cache limitations (and because of function call 
overhead---particularly since the functions are often virtual, in the C++ 
sense) it is much more efficient to enforce the boundary conditions on large
contiguous arrays of data. The same is true for
\code{DifferenceBoundaryConditions()}: it's a good idea to stack the requests 
for the distance to the closest image of an atom and compute then all at once. 

\subsubsection{Alternative Choices} 

A frequently used technique in MD when using periodic boundary conditions is
to stored ``scaled'' positions which take values between zero and one in each
direction. This allows certain tricks to be used in applying periodic boundary
conditions, such as adding and subtracting a ``magic'' number which has the
effect of putting a value outside the interval $(0,1)$ back into it in the 
appropriate way\footnote{This is processor dependent.}. This avoids the need 
for ``if'' statements which in principle are detrimental to performance. It has
 the disadvantage that positions must be
re-scaled in order to compute energies and forces. In our experience the
performance difference has been negligible, and we feel that using real space
coordinates is simpler and more intuitive.
 
To use the most general kinds of boundary conditions, one may have to add
some responsibilities to the \code{BoundaryConditions} component. For example 
one can
think of some complicated boundary conditions for which the two positions must
be provided to get their separation vector (not just the distance between
them). Another example is reflecting boundary conditions, where the
{\em velocities} of the atoms should be changed (note however that this kind of
\code{BoundaryConditions} could also be implemented as a \code{Constraint}).

\subsection{NeighborLocator}\label{neighborlocator}




\begin{figure}[thb]
\begin{center}
\epsfig{file=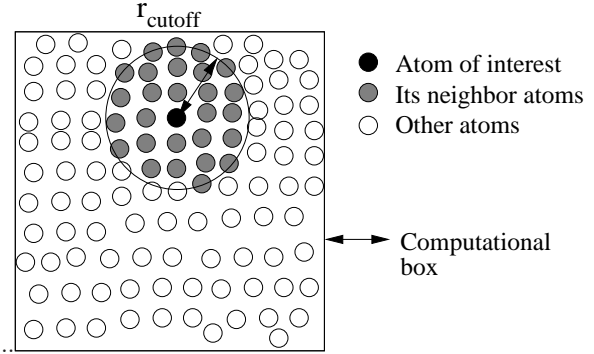, width = 3 in.}
\end{center}
\caption{\label{NLIllus}The main task of a NeighborLocator is to identify all 
neighbor-atoms (gray) for a given atom (black), given a cutoff distance 
$r_{\textrm{cutoff}}$}
\end{figure}

\begin{table}\label{NeighborLocatorMethods}
\begin{center}
\caption{Methods for \code{NeighborLocator}.}
\begin{tabular}{|l|}
\hline
NeighborLocator \\
\hline
SetPositionArray() \\
Neighbors() \\
HalfNeighbors() \\
UpdatePosition() \\
SetBoundaryConditions() \\
GetCutoffDistance() \\
SetCutoffDistance() \\
GetDrift() \\
SetDrift() \\
NeighborLocator *Copy() \\
DataIsValid() \\
UpdateMyData() \\
DisplayStatistics() \\
\hline
\end{tabular}
\end{center}
\end{table}

\subsubsection{Responsibilities}
It is very natural to assign the task of identifying which atoms are
located within some cutoff distance of a given atom to an individual component:
a \code{NeighborLocator}. Fig.~\ref{NLIllus} illustrates the general task.

It is possible to set clear responsibilities to a \code{NeighborLocator} and to
decouple it almost totally from the other components. In principle it doesn't
need to know anything about the atoms themselves, their possible constraints, 
or the details of their interaction: we have a well defined geometrical 
problem, and all the input that is needed is
\begin{enumerate}
\item A collection of points in space (the atoms-/-molecules location),
\item A cutoff distance,
\item The boundary conditions.
\end{enumerate}
With this input, the \code{NeighborLocator} must primarily be able to return 
all the
neighbors of a given atom. In addition, because a force calculation
uses Newton's third law, the \code{NeighborLocator} can be requested to return
 only ``half'' of the neighbors: when looped over all atoms, this 
\code{HalfNeighbors()}
function returns all the bonds exactly once (for example, for atom $i$,
\code{HalfNeighbors()} may return all the neighboring atoms with index $j>i$). 
This is particularly useful for pair potentials.

As a secondary responsibility, we found it extremely useful
if the \code{Neigh\-bor\-Loc\-at\-or} is able to return the index of all the 
atoms
which are located within a given distance of a given point (not
necessarily an atom). It is useful because it allows e.g. extensions of a
\code{NeighborLocator} to deal with type-dependent cutoffs (see below) and it 
is a natural responsibility since a \code{NeighborLocator} usually stores the
information that is needed to answer this question (like a cell list).

Finally, for efficiency reasons as well as for flexibility, we also included in
the \code{NeighborLocator}'s tasks the possibility of returning only the 
neighbors of a certain type. 

\subsubsection{Examples}

The most trivial example of \code{NeighborLocator} is the one that tests each 
time all the atoms whether their separation is shorter than the cutoff 
distance. A force
calculation using this type of ``\code{SimpleNeighborLocator}'' would be 
$O(N^2)$. It
has the advantage that it certainly works under any circumstances. In addition
having such a simple component is handy when the number of atoms we have to
work with is small.

In the opposite case, however, more clever methods must be implemented. The
principles of such tricks like Verlet neighbor lists and cell lists are well
documented in textbooks\cite{Allen/Tildesley:1987}. However making them as 
efficient as possible and decoupled enough from the other components is 
oftentimes delicate (see below).

Separating a \code{NeighborLocator} component from the others is extremely
 useful
because it can be used for far more than just the regular force calculation
between atoms. One could use it for other tasks, like the localization and
visualization of crystalline defects based on the coordination. A
\code{NeighborLocator} could also perform more sophisticated tasks. For example
we may want to break a subset of the atomic bonds (e.g. remove atoms from the 
neighbor list which lie across a half-plane in order to open a crack from a 
crystal). This could be done by a specially
designed \code{NeighborLocator} that would post-process the calculations of any
\code{NeighborLocator} (it would ``decorate'' another \code{NeighborLocator}, 
in the Design Pattern's language); there would be no need to dig in, or rewrite
 the \code{Potential}.


\subsubsection{Implementation, Efficiency, Flexibility} 

The \code{NeighborLocator} is primarily used by the \code{Potential} to 
calculate
forces and energies. It is a means to make the force calculation more
efficient by avoiding unnecessary computation. In addition, the
\code{Potential} generally must also compute the vector separating a pair of
interacting atoms, as well as their separating distance. However these
quantities are also computed by the \code{NeighborLocator} so it is very
natural that an inquiry for the neighbors of an atom, say $i$ returns at least:
\begin{enumerate}
\item The index of the neighboring atoms,
\item The separation vector between atom $i$ and its neighbors,
\item The squared distance between atom $i$ and its neighbors.
\end{enumerate}

The \code{NeighborLocator} generally stores some internal data. Depending on 
the concrete type of \code{NeighborLocator}, this could be a list of neighbors
 for each atom, or a cell list (a region enclosing the atoms is decomposed into
 cells, associated with each of which is a list of the atoms it contains). This
 data may become
invalid after the atoms have moved too much; at this time, the 
\code{NeighborLocator} must rebuild its data. A nice way to make sure that the
 data is always up to date is to implement a Subject-Observer-kind of
 relationship between the atomic positions and the \code{NeighborLocator}: each
 time the positions are changed a signal
is sent to the \code{NeighborLocator} that its ``subject'' was modified and 
that it must check whether its internal data is still accurate. This signal 
typically is an ``Update'' function; a \code{SimpleNeighborLocator} would do
 nothing, but a \code{NeighborList} would compute the maximum atomic 
displacement since the last building of the neighbor list and decide whether of
 not the list should be rebuilt.

As we said many times earlier the overall goal is to make the components of 
our MD program as independent as possible and concrete instances of a component
as interchangeable as possible. This means in particular that a 
\code{NeighborLocator}
should work without the knowledge of which kind of boundary conditions are in
effect (of course the associated overhead must be acceptable). If we want to
build a cell list (this is also true for a neighbor list, because to make a
neighbor list, it is more efficient to use a cell list), one must be careful to
correctly identify which cells are within the ``sphere of influence'' of
another. In fact this question is so closely related to the original one, that
a general method is to make no assumption about which cells are neighbors and
delegate the determination of neighboring cells to another ``upper''
\code{NeighborLocator} with a suitable cutoff. The centers of the non-empty
 cells
(squares/cubes or rectangles) are given to the upper \code{NeighborLocator}, 
with a cutoff $c_\mathrm{up}$ 

\begin{equation}
c_\mathrm{up} = c + \mathrm{Diagonal\ of\ the\ cells},
\end{equation}

\nod where $c$ is the original cutoff. This way one builds a whole hierarchy of
\code{NeighborLocator}s until the number of cells is small enough that a
\code{Sim\-ple\-Neigh\-bor\-Loc\-at\-or} can be effectively used. The nice 
aspect of this 
approach is that it works with any kind of boundary conditions, even if images
 (via the boundary conditions) of cells overlap.

\begin{figure}[thb]
\begin{center}
\epsfig{file=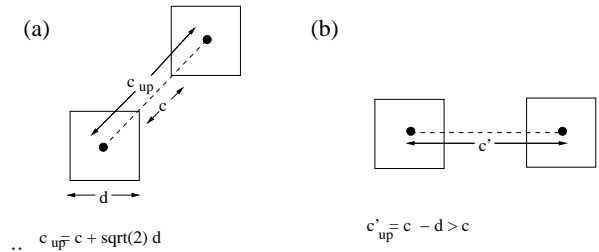, width = 3 in.}
\end{center}
\caption{\label{neighborCells}(a) Definition of the upper-level cutoff
 $c_\mathrm{up}$. This configuration indicates the farthest that two cells can
 be from each other and having an atom in one which is a neighbor of an atom i
n the other. (b) For this configuration, the cells' centers are separated by
 $c_\mathrm{up}$, but no atoms of one cell can be neighbors of atoms of the 
other. }
\end{figure}

Note that even if two cells are separated by a distance less than the cutoff
$c_\mathrm{up}$, depending on their relative position they may not necessarily
``relevant neighbor cells'' (being ``relevant'' meaning that atoms in one could
 possibly be neighbors of atoms in the other), as Fig.~\ref{neighborCells} (b)
 indicates. Thus in principle a gain in efficiency could be achieved by 
eliminating these cases, using some criterion based on vector between the 
center of one cell and the corners of the other.

\subsubsection{Alternative Choices}

The most efficient \code{Neighbor\-Locator} is called \code{Cell\-Neighbor\-List}, 
since it uses a cell list to construct neighbor-lists which are then used to
 provide neighbor 
information. Neighbor-lists have a large number of \code{int}s or pointers per
 atom 
and tend to dominate the memory requirements. It was thought that
 \code{CellNeighborLocator}, a neighbor-locator based on cell lists alone,
 would be quite useful for larger systems, but while it does take dramatically
 less 
memory, there are so many candidates for being neighbors that it is too slow.
 The problem is that one needs not only to search the current cell (say, with a
 cell length equal to the cutoff distance) but 26 other cells surrounding it. 
So, instead of  $\frac{4}{3}\pi r^3$, the volume to be searched is $27r^3$, a 
ratio of almost 6.5. 

There are still unresolved problems in the case of time-dependent boundary 
conditions, for example when a periodic supercell is sheared. This can cause
 unnecessary rebuilds of the neighbor-lists. Ideally one should allow for the 
shearing of the old positions in computing the maximum distance moved.

\subsection{Potential}\label{potential} 

This is the modularization that is most likely to be already implemented in
existing codes: most people want to have the freedom to change
potentials. Different potentials represent different materials, so the number 
of \code{Potential} classes one implements corresponds to the number of 
materials one wishes to simulate (of course, some potential classes will allow 
the simulation of multiple materials by varying the potential parameters).

\begin{table}\label{PotentialMethods}
\begin{center}
\caption{Methods for \code{Potential}.}
\begin{tabular}{|l|}
\hline
Potential  \\
\hline
CalculateEnergy() \\
CalculateForces() \\
CalculateEnergyAndForces() \\
CalculateStressTensor() \\
GetLengthScale() \\
GetEnergyCurvatureScale() \\
CalculatePairForceEnergy() \\
AtomicForcesEnergy() \\
CalculateForcesBetweenTypes() \\
CalculateAtomicEnergy() \\
CalculateHessianMatrix() \\
GetCutoffDistance2() \\
\hline
\end{tabular}
\end{center}
\end{table}

\subsubsection{Responsibilities} 

The primary responsibilities of the \code{Potential} class are two. Given a
\code{List\-Of\-Atoms} object and whatever arrays are necessary:

\begin{enumerate}
\item It calculates the current forces on the atoms, putting them into the 
passed array.
\item It calculates the potential energy of the current configuration, 
returning it as a double.
\end{enumerate}

Other methods that a \code{Potential} may have include one to calculate both 
energy and forces together (there are cases where both are needed and where it
 is much faster to calculate them together), to calculate the atomic energy 
(for a given atom), the atomic stress or the Hessian matrix. Not every function
 needs to be implemented; they are implemented in the base class as functions
 which do nothing except throw an exception. Thus client code may test the 
potential to see if it has the function. However as a minimum, a new potential
 should include \code{CalculateForces()} and \code{CalculateEnergy()}. 

\subsubsection{Examples} 

We currently have two versions of Lennard-Jones pair potentials, one our own, 
and one by Holian
et al.\cite{Holian:1991}. They differ in how their range is truncated 
spatially. We have the Stillinger-Weber\cite{Stillinger/Weber:1985}
potential for silicon, which includes three-body terms as well as the
EDIP\cite{Bazant/others:1997,Justo/others:1998} Si potential which is a many 
body potential. We also have the
Effective Medium Theory (EMT)\cite{Jacobsen/Stoltze/Norskov:1996} potential for
some fcc metals, with our implementation
specifically including Al, Cu, Ag, Au, Ni, Pd and Pt, and recently added
Baskes's Modified Embedded Atom Method (MEAM) potential\cite{Baskes:1992}, with
 parameters for 26 elements (pure elements only so far).

\subsubsection{Implementation, Efficiency, Flexibility} 

When say, the \code{CalculateForces()} function is called, the potential is 
passed the \code{ListOfAtoms} and an array for the forces. It gets a pointer to
the \code{Neigh\-bor\-Locator} of the atoms. If it is a pair potential, it 
loops 
over all atoms, and for each one calls \code{HalfNeighbors()} on the 
\code{NeighborLocator}, which returns the neighbors $j$ of atom $i$ with $j>i$ 
to avoid double counting. The \code{NeighborLocator} also returns the relative 
displacements (vectors and squared lengths) of the neighbors of atom $i$, which
 the \code{Potential} object
uses to compute the corresponding pair contributions to the forces or the 
energy. In the case of the forces these are added to the force array for atom
$i$, and their negatives to the force of each neighbor. For non-pair 
potentials, the looping is done differently but the
interaction with the \code{NeighborLocator} is similar (in some cases one calls
\code{Neighbors()} rather than \code{HalfNeighbors()}, e.g. for the three body
 terms of Stillinger-Weber). 

For extra efficiency, the interface to the \code{Neighbor\-Locator} has been
 designed so
that rather than computing all the force or energy contributions involving atom
$i$ before going onto the next atom in this loop, one can continue to fill the
arrays (of displacements and squares of displacements) with neighbors of
successive atoms $i$ in the loop, until some pre-determined size limit on the
array has been reached (as indicated by the return value of the 
\code{NeighborLocator}
function which is boolean). Then if the potential involves only floating point
operations these can be done faster when the data is packed into fewer, larger
arrays. Our EMT potential works this way. On the other hand this will not help
 if the potential involves secondary cutoffs
within the main cutoff, as in for example our \code{CutLennardJonesPotential}. 
Here the potential takes the usual form within the inner cutoff but has a 
different form between the inner and outer cutoffs. Because of this ``if'' 
statements are necessary, and so the operations are not all floating-point.  

Another efficiency point is that when possible the parameters of each potential
class are declared as constant variables, thus the compiler is allowed to make
optimizations that it might not otherwise make. This is not possible in
potentials such as EMT and MEAM where different choices of parameters are
allowed. 


\subsection{AtomsMover}\label{mover}


The algorithms associated with time evolution of the \code{ListOfAtoms} are
encapsulated as \code{AtomsMover}s. Every class of this family has a function
called Move which uses the potential to evolve the \code{ListOfAtoms} according
to the appropriate algorithm some number of time steps of some length. The Move
function can be considered a transformation of the \code{ListOfAtoms}, but of a
particular type---one associated with time steps and potentials, what we might
refer to as ``dynamics''.

\begin{table}\label{AtomsMoverMethods}
\begin{center}
\caption{Methods for \code{AtomsMover}}
\begin{tabular}{|l|l|l|}
\hline
AtomsMover \\
\hline
Move()       \\
SetPotential()\\
GetDt()       \\
SetDt()       \\
GetTime()     \\
SetTime()     \\
\hline
\end{tabular}
\end{center}
\end{table}

\subsubsection{Responsibilities} 

The responsibilities of a \code{Mover} are less well defined than those of 
components such as \code{Neighbor\-Locator} and \code{Potential}. Particularly in
 the case of the \code{Neighbor\-Locator}, the information returned should be 
independent of which implementation of a \code{NeighborLocator} one has used. 
For a potential, of course the forces and energy will vary from one potential 
to another but the meaning of \code{Calculate\-Forces()} and
 \code{Calculate\-Energy()}
 is the same for all (for instance the forces are always the negative gradient 
of the energy). For movers, there are  not such specific requirements. Several 
movers implement time stepping 
algorithms, but it is not required that these give identical results for a 
given \code{ListOfAtoms}. One important mover, \code{Quickmin}, actually 
performs energy minimization rather than physical time evolution. It is 
included with the movers---unlike our alternative minimization class, a 
general purpose conjugate gradients algorithm---because the algorithm is 
closely related to molecular dynamics algorithms. All \code{Mover}s are linked
 to a \code{Potential} object, and store a
value of time step, and number of steps to perform. The time steps performed in
 a single call to Move are
what are sometimes called the {\em minor} time steps. Each call to move, made
from some outer loop, constitutes a major time step. 

\subsubsection{Examples} 

Our \code{Mover}s include a slightly non-standard Verlet time-stepping
 algorithm, as 
well as the Gear
Predictor-Corrector algorithm. For thermalized time-stepping there is a
\code{LangevinAtomsMover} (which implements the Langevin equation with a given
temperature and friction), a \code{HooverAtomsMover}\cite{Holian/others:1990}, 
a \code{VerletThermalizeAtomsMover} which randomizes velocities before 
performing the
time steps. We also implement the {\em QuickMin} algorithm for minimizing a
molecular dynamics system as a \code{Mover} since it is closely related to the
 Verlet
algorithm. To implement Conjugate-Gradients (CG) minimization we provide an 
interface class (Mediator Design Pattern) which allows a general purpose CG
class to operate on a given \code{ListOfAtoms} using a given \code{Potential}.

\subsubsection{Implementation, Efficiency, Flexibility} 

For movers, the implementation is for the most part as straightforward as
writing out the formulas for the algorithm in terms of functions on 
\code{ListOfAtoms}
and potential, taking care only to do use the array versions of operations on
the \code{ListOfAtoms}---one should never write a {\tt for} loop in which the
positions are updated one by one. Apart from the loop overhead, each call to
\code{ListOfAtoms::SetCartesianPosition()} or \code{ListOfAtoms::IncrementCartesianPosition()}
 (no \textit{s}) entails
 a call to the \code{NeighborLocator} to update its internal variables, which 
will possibly involve initiating communication between processors in a parallel
simulation (although this process could be postponed until actually
called for). Clearly this is bad.

Thus the loop for \code{Ver\-let\-At\-oms\-Mov\-er} is not
 much more than a call to \code{Cal\-cul\-ate\-For\-ces()} (\code{Potential}) 
followed by
 a call to \code{Inc\-re\-ment\-Cart\-es\-ian\-Vel\-oc\-it\-ies()} and a call 
to 
\code{Inc\-re\-ment\-Car\-tes\-ian\-Pos\-it\-ions()} (both 
\code{List\-Of\-At\-oms}). At the
 beginning and end of the loops there are increments by half a time step to
correctly implement the Verlet
algorithm and have the velocities and positions correct and consistent when the
function exits. Extra details concern the handling of constraints (see section
\ref{constraint}).

\subsection{Transformer}\label{transformer}
 For all other
transformations on the \code{ListOfAtoms} we have another family of components 
called \code{Transformer}s which have a similar interface, with the function 
being called Transform in this case.

\subsubsection{Resposibilities}
The responsibilities of a \code{Transformer} are none at all, other than to
 provide a function called Transform, to which is passed a pointer to a
 \code{ListOfAtoms}.

\begin{figure}[thb]
\begin{center}
\epsfig{file=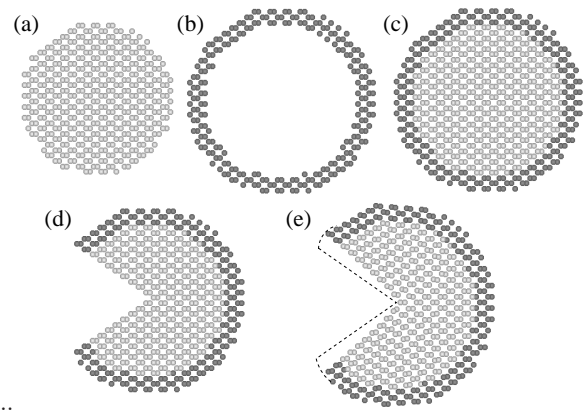, width = 3 in.}
\end{center}
\caption{\label{InitTransfIllus}Illustration of setting up a system by 
combining
\code{Initializer}s and \code{Transformer}s. First lists of atoms occupying
a disk-region (a) and an annular region (b) are created. These are combined as 
branches of the main \code{ListOfAtoms} (c). Next a \code{NotchMaker} (a 
\code{Transformer}) is used to remove the appropriate atoms so as to make a 
notch of specified center and opening angle (d). Finally an example of an 
\code{ElasticFieldTransformer}, namely, a \code{NotchFieldDisplacer}, is used
to displace the atoms into an initial loaded state by applying the linear 
elastic displacement appropriate for a notched body (e).}
\end{figure}

\begin{table}\label{TransformerMethods}
\begin{center}
\caption{Methods for \code{Transformer} and 
\code{Elastic\-Field\-Transformer}.}
\begin{tabular}{|l|l|l|}
\hline
Transformer & ElasticFieldTransformer\\
\hline
Transform() & ElasticDisplacement() \\
            & GradientWRTPosition() \\
\hline
\end{tabular}
\end{center}
\end{table}

\subsubsection{Examples}

Since \code{Trans\-for\-mer} is so general, we have many examples. Some, such 
as \code{Void\-Mak\-er}, \code{Notch\-Mak\-er} and \code{Ov\-er\-lap\-Prun\-er}
 cut away parts of a \code{List\-Of\-At\-oms} defined
by some geometrical criterion. These are frequently used in conjunction with an
Initializer, where one first creates a simple shape filled with atoms and then
cuts away pieces to achieve the actual desired geometry (see 
Fig.~\ref{InitTransfIllus}). A subclass of
\code{Trans\-for\-mer} is \code{El\-as\-tic\-Field\-Trans\-for\-mer}, which 
covers \code{Trans\-for\-mer}s whose
transformation is associated with a displacement field as in continuum
elasticity. In these the \code{Transform()} function is passed to a separate 
function called \code{El\-as\-tic\-Field()}. This is useful in cases where one
 wants to be able to
evaluate the \code{ElasticField()} function without actually transforming any
atoms. Also, the \code{EulerCoordinateTransformer} class keeps a list of
\code{Elas\-tic\-Field\-Trans\-for\-mer}s and iteratively determines using 
their \code{ElasticField()} functions, the resultant displacement given by the 
elastic field formula interpreted as an expression in Eulerian coordinates. 
Examples of \code{Elas\-tic\-Field\-Trans\-for\-mer}s include \code{(An\-iso\-trop\-ic)Dis\-loc\-ation\-Mak\-er}, which
implements the standard displacement formulas for straight edge and screw
dislocations in (an)isotropic elasticity theory; 
\code{Notch\-Field\-Disp\-lac\-er}, which
implements the displacement field for a notched or cracked sample, and
\code{(An)\-iso\-trop\-ic\-Mul\-ti\-pole\-Field} which supplies the entire set
 of terms in the
general solution for quasi-two-dimensional elastic theory in a circular
geometry, apart from the dislocation (log) terms, and the terms which grow with
distance from the origin. 

\subsubsection{Implementation, Efficiency, Flexibility}
 For transformers of course there is little one can say in
general about implementation, but the point about using array operations to
 change atomic positions is just as valid. So in the case of applying an
 elastic displacement for example, the increments should be computed as a
 separate array which is then added to the positions using
 \code{Inc\-re\-ment\-Car\-tes\-ian\-Pos\-it\-ions()}.

\subsection{Constraint}\label{constraint}
 

The simplest MD application uses periodic boundary conditions (PBC). There are
tricks one can use to allow application of stress or strain to the system using
PBC, but there are often cases, particularly when an atomistic simulation is
coupled to a larger length scale simulation, when more physical boundary
conditions are necessary, such as a layer of atoms which is fixed, or moves
uniformly, or has a constant force on it. To effect such behavior we add a {\em
Constraint} to the appropriate branch of the \code{ListOfAtoms}. While in some
 numerical methods, such as the Finite Element Method, constraints are directly
 incorporated via problem formulation into the set of update equations to be 
solved, in MD, they are often applied after updated positions and velocities
have been computed. To allow for both possibilities we provide two different
interfaces for Constraint classes.

\begin{table}\label{ConstraintMethods}
\begin{center}
\caption{Methods for \code{Constraint}.}
\begin{tabular}{|l|}
\hline
Constraint  \\
\hline
AdjustPositions() \\
AdjustVelocities() \\
AdjustForces() \\
AdjustForceIncrements() \\
NumberGeneralizedCoordinates() \\
SetGeneralizedCoordinates() \\
GetGeneralizedCoordinates() \\
IncrementGeneralizedCoordinate() \\
CalculateGeneralizedForces() \\
DerivativesWRTGeneralizedCoordinates() \\
\hline
\end{tabular}
\end{center}
\end{table}

\subsubsection{Responsibilities} 

The two interfaces are not
both implemented for all existing \code{Constraint}s, because the nature of 
some constraints makes it difficult to implement one or the other; in some case
it is even ambiguous how exactly a given interface should behave for a
 particular constraint (e.g., the generalized coordinates for a 
\code{FixedCenterOfMassConstraint}, where one would have to make an arbitrary
 choice in order to define the generalized coordinates). 

\begin{description}
\item[The ``Adjust'' Interface] This is the most commonly used method of
incorporating the constraints. \code{Constraint} classes provide methods to
 Adjust
position, velocity and force arrays. As long as these functions are called (for
example by a mover) any time that the positions and velocities are incremented
or set, or any time that the forces are calculated, the appropriate Adjust
function will ensure that the positions satisfy the constraint, and the
velocities and forces are such that the positions continue to satisfy the
constraint when the velocities or forces are used to increment them. Some
constraints deal explicitly with velocities and forces rather than positions
(e.g. \code{UniformlyMovingBody} and \code{ExternalForceConstraint}) . These 
are not
constraints in the usual mechanics sense, although if the velocities (or
momenta) are considered on an equal footing with positions, then a velocity
constraint has an equivalent status to a position constraint. A constraint on
forces---our example involves the addition of additional force, as in an {\em
external} force---is strictly a change of the potential, but it is
more easily implemented as a constraint using the Adjust interface. 

\item[The Generalized Coordinates Interface] This interface is less well
developed; it more explicitly corresponds with the classical notion of a
constraint on dynamical degrees of freedom (DOFs). To use this interface means
rather than using the standard interface with the \code{ListOfAtoms}
(\code{Inc\-re\-ment\-Cart\-es\-ian\-Pos\-it\-ions()}, 
\code{Inc\-re\-ment\-Car\-tes\-ian\-Vel\-oc\-it\-ies()} etc)
 and calling instead 
\code{Inc\-re\-ment\-Gen\-er\-al\-ized\-Co\-or\-din\-ate()} and 
\code{Cal\-cu\-late\-Gen\-er\-al\-ized\-For\-ces()}. At this
time there is no method to access the generalized velocities, which if one was
properly using generalized coordinates would be stored instead of, or in
addition to actual velocities (which would be slaved to the generalized
velocities). 
\end{description}

\subsubsection{Examples} 

The base class \code{Constraint} is itself a valid constraint---it is the null 
constraint, in that its Adjust functions do nothing. A \code{ListOfAtoms} is 
born
having its \code{Constraint} object be a base class instance. To set a new 
constraint, the new constraint is created externally, then passed in a call to
\code{List\-Of\-Atoms::Set\-Con\-straint()}, which deletes whichever 
constraint was 
previously there, and reassigns the LOA's pointer.

The most frequently used \code{Constraint}\footnote{in our experience to date} 
is \code{FixedBody}. When this is attached to a \code{ListOfAtoms}, subsequent 
calls to change
the positions will have no effect. The Adjust interface simply stores the
initial positions and whenever the positions are changed and Adjust is called,
they are changed right back. The Generalized Coordinate interface exists in 
that it returns zero as the number of independent DOFs and thus allows no
changes to the positions. 

For simulations which involve applying traction to the surface of a sample, 
one separates the outer few layers of atoms into a separate branch of the  
\code{ListOfAtoms} tree, and adds an \code{ExternalForceConstraint} to that 
branch. Only the
\code{AdjustForces()} method does
anything; it adds a given external force divided by the number of atoms in the
branch to each atom in the branch. A weakness with this is that if one called
\code{AdjustForces()} more than once per call to \code{CalculateForces()} from the 
\code{Potential}, 
the actual extra force would be several times what was intended. Ideally an 
Adjust function should be idempotent---calling it several times should have 
the same effect as calling it once.



\subsubsection{Implementation, Efficiency, Flexibility} 

\code{Mover}s do not communicate with the constraint directly. In the case of
\code{Ad\-just\-Pos\-it\-ions()}, nothing extra needs to be done because this 
is handled 
by the \code{Set\-Car\-tes\-ian\-Pos\-it\-ions()} and 
\code{Inc\-re\-ment\-Car\-tes\-ian\-Pos\-it\-ions()} methods. These call
an \code{Ad\-just\-Pos\-it\-ions()} method of the \code{List\-Of\-At\-oms}, 
which does two things: calls the corresponding method on its own 
\code{Cons\-traint}, and calls the corresponding 
method on its branches. \code{Inc\-re\-ment\-Vel\-oc\-it\-ies()} works the same
 way. \code{Ad\-just\-For\-ces()} is 
called by \code{Mov\-er}s, since it is they that own force 
arrays.
 Again first the method of the LOA's own \code{Cons\-traint} is called
on the whole list, then the same method on each branch of the LOA is called on
the corresponding branch of the force array.

We have to add an additional \code{Adjust()} function to allow certain movers to
behave properly: \code{Adjust\-Force\-Increments()}. This is applied to for example
the random amounts that are added to the forces by \code{Langevin\-Atoms\-Mover}, in
order to zero out or average certain components of the force array. It is mostly
identical to \code{AdjustForces()}, except for \code{ExternalForceConstraint}
where it is important to not add the external forces---since these are
separately added to the force array, and must only be added once!

\subsubsection{Alternative Choices} 

The \code{Constraint} component is very incomplete. It is easy to imagine 
constraints which cannot easily be implemented using either interface and 
movers which 
would break the present mechanism for enforcing constraints. Also for each 
currently existing constraint one can imagine possibly more efficient 
alternative implementations. A case where the present mechanism is 
demonstrably weak is the \code{FixedBody} constraint. The \code{AdjustForces()} 
function of this constraint zeros the forces for a \code{ListOfAtoms} with this
 constraint. This is clearly necessary if one is using 
\code{LangevinAtomsmover} because the algorithm for Langevin dynamics has a 
step which requires incrementing the positions by 
an amount proportional to the forces. However in some applications one might 
want to know the forces on
those atoms for reasons other than to move them---these being the reaction
forces required to hold these atoms in place, which are often of interest. In
particular, sometimes the atoms in the \code{FixedBody} {\it do} move; they are
not moved my an MD algorithm, but rather by an external object, such as an
\code{ElasticFieldTransformer} whose parameters are being evolved by some high 
level algorithm. The high level algorithm might need to calculate forces on 
those parameters, which forces are a function of the atomic forces. To avoid 
zeroing 
the forces one might make a separate call to calculate forces and not call
\code{AdjustForces()}. However this might result in unnecessary repetition of
 the
force calculation. Furthermore if the high level procedure is taking place in
Python, it may not be possible to control whether the constraint is applied or 
not---the current interface to \code{CalculateForces()} for all \code{Potential} 
objects, from Python, includes a call to \code{AdjustForces}. To get around 
this one needs to remove the \code{FixedBody} Constraint (which is done in 
practice by replacing it with a base class \code{Constraint} object). However 
this requires the user to be more aware of the details of what 
\code{Constraint}s are attached to what \code{ListOfAtoms} objects than is 
desirable, although maybe not more aware than is necessary; that is to say, 
such specificity may be unavoidable.

It may be safer to give the constraints more responsibility instead of having
the post-processing step we currently use. So rather than call 
\code{Inc\-re\-ment\-Car\-tes\-ian\-Pos\-it\-ions()}
and then Adjust, pass the Increment call on to the constraint so that it can 
handle the entire process. This would speed up the \code{FixedBody} for 
example, because then the \code{Constraint} could simply not pass on the 
instruction---which
 saves time on incrementing and then resetting to the original positions. The
 same could be said for \code{UniformlyMovingBody}. 

For \code{ExternalForceConstraint}, it would be safer to pass the 
\code{Cal\-cul\-ate\-For\-ces()}
command to the constraint which would calculate the potential forces from
scratch and then the external force. Subsequent calls would repeat this rather
than accumulate many times the desired external force. 

Rather than having two interfaces within one class, it might
make sense to separate them into two different classes. This has been done in
an MD implementation known as CampOS/Asap~\cite{campos}, where the 
\code{GeneralizedCoordinate}
interface corresponds to a ``Filter'' class. Other objects, such as minimizers
 interact with the \code{ListOfAtoms} through a Filter object which connects 
the generalized coordinates to the actual atomic positions.

\subsection{ListOfAtomsObserver}\label{listofatomsobserver}


A key innovation of our work is the separation of core computation from
measurement. The code for these is often found together but it does not need to
be. One should be able to code an \code{AtomsMover} without thinking about what
measurements are to be made on the system during a simulation run. To implement
this separation we have used the so-called {\it Subject-Observer} Design
Pattern \cite{Gamma/others:1995}. For the purpose of this pattern, a {\it 
measurement} is any function or operation that looks at the simulation's 
(\code{ListOfAtoms}'s) data
but does not modify it. This includes statistical averaging of various kinds,
graphics and visualization, saving simulation states to disk and others.

\begin{table}\label{ListOfAtomsObserverMethods}
\begin{center}
\caption{Methods for \code{ListOfAtomsObserver}.}
\begin{tabular}{|l|}
\hline
ListOfAtomsObserver  \\
\hline
Update() \\
GetNotifyLevel() \\
\hline
\end{tabular}
\end{center}
\end{table}

\subsubsection{Responsibilities} 

The Subject-Observer pattern designates one class, in our case 
\code{ListOfAtoms}, as a {\it subject}. As a subject, \code{ListOfAtoms} has a
 function \code{AddObserver()}, to which a \code{ListOfAtomsObserver} (referred
 to as Observer for brevity) object is 
passed. A pointer to it is then kept on a list by the \code{ListOfAtoms}. The
 other part of the
pattern on the \code{ListOfAtoms} side is a function called Notify. A subclass
 of  Observer must provide an Update function. When Notify is called on the
\code{ListOfAtoms}, it iterates over the list of attached Observers calling
 each  \code{Update()} function in turn, passing itself (the \code{ListOfAtoms})
 as an argument. The observer can then access data of the \code{ListOfAtoms}
 and perform its task. It may not alter the \code{ListOfAtoms}.

\subsubsection{Implementation, Efficiency, Flexibility} 

The implementation of the Subject-Observer pattern in the Digital Material
follow Ref. \cite{Gamma/others:1995} quite closely. The Notify function is 
generally
called at the outer level as part of the outer loop. Thus suppose one wants to
run a simulation for 10000 time steps, recording the potential and kinetic
energy every 100 steps. One first would attach an \code{EnergyObserver} to the
\code{ListOfAtoms}, set the number of steps in the \code{AtomsMover} to
be 100, and then have an outer loop with 100 iterations, in which one calls
first the Move function of the \code{Mover} and then \code{Notify()} on the 
\code{ListOfAtoms}.

An enhancement to the standard Observer pattern that we have implemented is to
associate an integer, called the {\it notify level}, to each observer as it is
created. The default value of the notify level is 0. The \code{Notify()} 
function of \code{ListOfAtoms} now takes an integer argument, called 
\code{notifyLevel}, also with a
default value of zero. For nonzero argument, only those observers whose own
notify levels are less than or equal to the notifyLevel argument of 
\code{Notify()},
are actually updated. For example, suppose we want to measure the energy for
the purposes of averaging (other some other function of the atomic state) every
 10 time steps, and save the atomic
configuration to a file every 100 time steps. These processes would be handled
by say an \code{EnergyObserver} and a FileObserver, respectively. We could 
create the  \code{EnergyObserver} with a notify level of 0 and the 
\code{FileObserver} with a notify
level of 1. In the main loop, having set the minor time step to be 10, when the
major time step is a multiple of 10 we call Notify with an argument of 1,
otherwise we call it with an argument of zero. Alternatively we could imagine 
a more complicated high level control structure with nesting of loops, the 
inner one(s) calling Notify with argument 0 and the outer one(s) with argument
unity. In general if an observer is not to be always updated, it should have a
notify level greater than zero.

\subsubsection{Examples} 

\begin{enumerate}
\item \code{EnergyObserver}: measures potential and kinetic energy and stores
 totals
for the purpose of statistical analysis of these quantities (mean, variance and
 related quantities).
\item \code{PlotAtomsObserver}: plots atoms using the \code{PlotAtoms} package.
\item \code{RasMolObserver}: same using the \code{RasMol} package
\item \code{CheckPointObserver}: a simple implementation of saving the state of
 a simulation to disk, using Serialization (see \ref{serialization} below).
\item \code{StressIntensityObserver}: for crack simulations, it computes an
 estimate of the stress intensity factor  around a crack.
\item \code{PythonLOAObserver}: this is very important as it allows new 
observers to
be defined purely within Python and be attached and Notified exactly as if they
were C++ observers. The Python observer creates an instance of this class,
passing a pointer to its own Update function.
\end{enumerate}


\section{\label{infrastructure}Infrastructure}

\subsection{Parallelization}\label{parallelization}

It is clear that any modern MD code must be parallelizable. Modern scientific
computation is relying more and more on clusters of processors, especially as
it becomes easier to build these from ``off-the-shelf'' hardware. Furthermore, 
MD as used for materials modeling lends itself very well to parallelization
since interactions are typically short-ranged, and the amount of computation 
between communication steps can be rather substantial. Thus if different 
processors
handle distinct regions of space, a given processor needs only to know about
the positions of atoms on ``neighboring'' processors which are within a cutoff
distance of itself. For a large enough number of atoms per processor, this is a
reasonably small fraction of a processor's atoms whose positions need to be
communicated to other processors each time step.

We have implemented parallelization in a way which is almost transparent to the
user. To make an application (either a \texttt{main} function or a Python 
script) run in parallel, typically only a few lines have to be added. These
involve creating parallel versions of \code{Potential}, \code{NeighborLocator}
 and \code{AtomsInitializer}, which wrap the serial versions of these---thus, 
it is still necessary to create the ordinary serial object. But then one passes
 its address to the parallel version and from then on refers to the parallel 
version (e.g. the \code{AtomsMover} is given the \code{ParallelPotential}, and 
the \code{ListOfAtoms} is given the 
\code{Par\-all\-el\-Neigh\-bor\-Loc\-at\-or}). To
 implement parallelization we need the following:

\begin{enumerate}
\item A means of defining which atoms belong to which processors
\item A means of distributing atoms among processors
\item A way for a processor to know the positions of atoms on other processors 
that it needs to properly compute forces on its own atoms
\item A way to redistribute atoms between processors when their positions have
changed sufficiently, which must also redistribute corresponding items in
related arrays (velocities, forces, etc.)
\end{enumerate}


\begin{figure}[thb]
\begin{center}
\epsfig{file=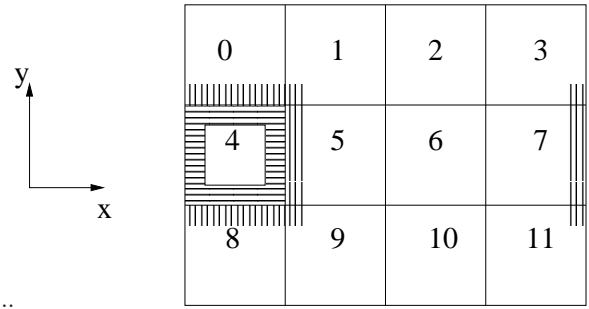, width = 3 in.}
\end{center}
\caption{\label{ghostAtomsIllus} Typical layout of 12 processors for a 2D 
simulation with periodic boundary conditions. The atoms belonging to processor
 4 which are ``interesting'' to 
other processors (their positions need to be communicated to them) are in the
region with horizontal shading. The regions with vertical shading contain atoms
on other processors whose positions are needed by processor 4. These correspond
 to processor 4's ``ghost atoms''.}
\end{figure}

\begin{table*}\label{ParallelizationMethods}
\begin{center}
\caption{Methods for \code{Migrator}, \code{Synchronizer}, 
\code{DomainSubdivision}.}
\begin{tabular}{|l|l|l|}
\hline
Migrator &  Synchronizer & DomainSubdivision \\
\hline
SetPositionsArray()  & SetPositionsArray   & Update() \\
Migrate()            & Rebuild()           & WhichDomain() \\
                     & Synchronize()       & PointNearDomain() \\
                     & GetGhostsShape()    & FindPointsNearDomain() \\
\hline
\end{tabular}
\end{center}
\end{table*}

\subsubsection{Ghost Atoms and Synchronization}

All the atoms of the system are distributed somehow among the processors, such
that each atom ``belongs'' to exactly one processor. The definition of 
``belongs'' is assigned to the \code{DomainSubdivision} abstract class, which 
so far has one subclass, \code{HomogeneousDomainSubdivision}, which divides a 
rectangular space equally into domains, and associates each domain with a
 processor. A  processor is responsible
for computing the forces on all of its own atoms, as well as updating 
velocities and positions. All processors are running the same program, and thus
work with \code{ListOfAtoms} objects of identical tree structure. However the 
number of atoms on each leaf will differ from processor to processor. The 
function \code{GetNumber()} will return just the number for that processor---to
get the total number of atoms, an \code{AllReduce()}-type operation must take
place. To calculate forces, a processor needs to know the positions of certain
 atoms belonging to
other processors. Such atoms are called ``ghost atoms'' from that processor's
point of view. The assignment of atoms to processors, and the idea of ghost
 atoms, are illustrated in Fig.~\ref{ghostAtomsIllus}. The process of 
obtaining information about ghost atoms (their 
amount, types and positions) is called \code{Synchronization}, and is handled
 by the
\code{Synchronizer} class. This information is accessed by a given 
\code{DMArray} through the \code{SynchronizationKit} class, which provides an 
extension of a \code{DMArray},
containing separate \code{DMArray}s \code{ghostAtoms} (which will actually 
contain the ghost atoms' data), and \code{ex\-ten\-ded\-Arr\-ay}, which exists 
to be a parent array for the given array and ghostAtoms. It is the 
\code{extendedArray} which eventually is seen by 
the serial \code{NeighborLocator} on each processor. \code{Synchronizer}'s main
 methods are 
(1) \code{Rebuild()}, which constructs for the current processor the lists of its own
atoms which are ``interesting'' to each other processor---it consults the
\code{Dom\-ain\-Sub\-div\-is\-ion} for the definition of 
``interesting''---and (2) 
\code{Synchronize()}, which communicates the positions of the interesting atoms
 to the
appropriate processor.

\subsubsection{Migration}

The process of transferring atoms between processors when their positions have
changed appropriately is called Migration, handled by the \code{Migrator} 
class. We
need to be careful about what it means to ``transfer atoms''. In some OOP
implementations of MD, there is an Atom class, which contains the position,
velocity, force, mass etc. As discussed in subsection~\ref{listofatoms} this is
not efficient when it comes to updating the positions or the velocities, but 
it would make more obvious what block
of data should be transferred to the other processor. We could say that what
ever arrays are in a given \code{ListOfAtoms} should be transferred, but that
 would leave the force arrays, which are managed not by the \code{ListOfAtoms},
 but by the \code{Atoms\-Mover}, unchanged. The mechanism we have designed to 
automatically handle the transfer of all atomic data between processors is 
called the  \textit{sibling mechanism}. \code{DMArray}s which are siblings of 
each other will all be migrated when one is---the one being the position array,
 typically. The information about what siblings an array has is stored in the
\code{SynchronizationToken} class. An array is made a sibling of a previously 
existing one if the old array is passed to the constructor of the new one. The 
\code{Migrate()} function of \code{Migrator} assembles data to be
communicated, from all its sibling \code{DMArray}s, in the communication 
buffers, makes an \code{AllToAll()} communication, transfers the newly received
data from other processors to its own \code{DMArray}s. Finally it calls the 
\code{Synchronizer}'s \code{Rebuild()} function to recreate the lists.

\subsubsection{For the user: parallelization wrapper classes}

A user writing a Python script or main function for a parallel simulation must
create instances of \code{Dom\-ain\-Sub\-div\-is\-ion}, \code{Synchronizer} and
\code{Mig\-ra\-tor}, and as well
of parallelization ``wrappers'' around the usual 
\code{Atoms\-In\-it\-ial\-iz\-er},
\code{Neigh\-bor\-Loc\-at\-or} and \code{Pot\-ent\-ial}. A wrapper is a 
subclass 
containing a pointer to an instance of one of the standard (serial) subclasses.
Most method are passed on to the serial class, possibly with additional
processing appropriate for a parallel simulation (e.g., summation of a return 
value over all processors).

\begin{description}
\item {\code{ParallelAtomsInitializer}:} The regular \code{Atoms\-Initializer}
(e.g. \code{Rect\-ang\-ul\-ar\-Clus\-ter\-In\-it\-ial\-iz\-er}) is passed to 
this as a constructor
argument. When Create() is called, the regular \code{Atoms-Initializer}'s 
Create() is called on processor zero only. On other processors, copies of the 
resulting \code{List\-Of\-Atoms} with the same tree structure, but empty leaves, 
are created. The first call to the \code{NeighborLocator} to update will lead 
to atoms being Migrated
for the first time. This happens before any neighbor lists are constructed, so 
processor zero will not need to provide the large amount of memory that storing
these for the whole system. However, the fact that it briefly stores the 
positions of all other atoms does put an eventual limit on the scalability; 
once the total number of atoms approaches $10^8$, the procedure would probably
have to be modified.
\item {\code{\-Par\-all\-el\-Neigh\-bor\-Loc\-at\-or}:} Wraps around the 
regular
\code{Neigh\-bor\-Loc\-at\-or}. Its \code{Up\-date\-My\-Data()} function calls 
the \code{Migrate()} and \code{Syn\-chron\-ize()} as well as the
regular \code{Neigh\-bor\-Loc\-at\-or}'s \code{UpdateMyData()}. Also, its 
\code{Set\-Pos\-it\-ion\-Arr\-ay()} causes arrays to be allocated for ghost 
atoms'
 positions, calls \code{Migrate()} and
\code{Synchronize()}, and calls the regular \code{Neigh\-bor\-Loc\-at\-or}'s 
\code{Set\-Pos\-it\-ion\-Arr\-ay()}, passing it the 
\code{ex\-ten\-ded\-Arr\-ay} (i.e. including the ghost positions) obtained
from the \code{Syn\-chron\-iz\-at\-ion\-Kit}.
\item {\code{Par\-all\-el\-Pot\-ent\-ial}:} Wraps around the regular potential.
 The only extra tasks it performs are to sum the energy from all processors, 
and to ask the \code{Syn\-chron\-iz\-at\-ion\-Kit} to allocate the ghost-atom
 arrays. In the case
 that there are intermediate arrays during force and energy computation, 
however, these will need to be synchronized by the regular potential, by 
appropriate calls to the \code{Synchronizer}. This is the case for many-body
 potentials like EMT, and 
must be kept in mind when writing a new many-body potential. 
\end{description}

\subsubsection{Interaction with MPI through DMProtocol}

We have used MPI to send messages between processors, but have added 
a layer of abstraction between Digital Material code and MPI, known as the 
Protocol, so that the use of MPI is not hard-wired into the code. The base 
class DMProtocol provides an interface with methods such as Broadcast(), 
\code{AllToAll()}, \code{GetNumberOfProcessors()},  
\code{GetProcessorNumber}(), etc. These are 
overloaded in subclasses \code{MPIProtocol}, which connects these methods to 
actual 
MPI calls, and \code{LocalProtocol}, which has trivial implementations of the 
above 
methods for use in serial operation (i.e., return 0 for processor number, 1 
for number of processors, do nothing for other methods, etc.). General
 DigitalMaterial
code only ever knows about a global pointer, dmProtocol, to the base class
DMProtocol. This points to the Protocol object currently in use. Creation of
the Protocol object is controlled by the class \code{ProtocolFactory} 
(Singleton pattern---only one object ever exists), whose \code{SetProtocol()} 
method allows different Protocols to be set. At present the only protocols are
 \code{MPIProtocol} and \code{LocalProtocol}, but the system could handle a
 different message-passing
system if one were available. When using Python for a parallel application, a
special version of the python executable must be used, known as 
mpipython\cite{mpipython}. 
This is necessary in order that \code{MPI\_Init()} be called before anything 
else. When the \code{ProtocolFactory} is instantiated (this happens statically)
 it checks if \code{MPI\_Init()} has been 
called; if so it sets the Protocol as MPI. This will be the case in a Python
application; the user need not do anything. In a pure C++ application, near 
the start of main() should be a 
call to \code{MPI\_Init()} followed by a call to the
\code{ProtocolFactory::SetProtocol()}, passing 
\code{``ProtocolFactory::World''} which is an
enumerated type representing \code{MPIProtocol}.

\subsubsection{Parallelization and new Digital Material classes}

For the researcher who wishes to write a new interatomic potential within the
DigitalMaterial framework, there is very little that needs to be kept
in mind for the purposes of parallelization, since the 
\code{ParallelNeighborLocator}
takes care of most details. The main thing is to be aware in force calculations
that some of the neighbors returned by the \code{NeighborLocator} will have 
atom numbers apparently too high (i.e., $j > nAtoms$), these being ghost atoms.
For such atoms no space exists in that processor's force array and a dummy 
variable should be used to hold their forces:

\begin{verbatim}
double dummyForce[DIMENSION];
...
if(j<nAtoms) forceJ = (*forces)[j];
    else forceJ = dummyForce;
\end{verbatim}

Incidentally, this is the reason that the \code{Half\-Neigh\-bors()} function 
of \code{Neigh\-bor\-Loc\-at\-or} returns neighbors $j$ with $j>i$, rather with
 $j<i$---it allows ghost atoms to be included.

In writing transformers, it is even more necessary to use array access to the
\code{List\-Of\-At\-oms}; that is, use 
\code{Set\-Car\-tes\-ian\-Pos\-it\-ions()} rather than 
looping and calling \code{Set\-Car\-tes\-ian\-Pos\-it\-ion()}, since the latter
 entails the \code{Neigh\-bor\-Loc\-at\-or} updating itself, and thus 
communication between 
processors, for each loop iteration (as mentioned before this could be avoided 
by having the \code{Neigh\-bor\-Loc\-at\-or} take note of updates but not 
initiate a 
rebuild until actually needed---that is, until a call to Neighbors() or 
\code{Half\-Neigh\-bors()} is made).

\subsubsection{Alternative choices}

We have not implemented any kind of dynamic load-balancing scheme. In solid
mechanics atoms do not tend to move a whole lot, nor does density tend to 
change much so that if the atoms are well distributed at the start of the
simulation they will more or less remain so. However if one wanted to
implement load balancing, which would amount to redefining processor
boundaries dynamically, one could implement a new subclass of 
\code{DomainSubdivision} which would implement whatever algorithm was to be 
used for the load-balancing.

\subsection{Serialization}\label{serialization}

Serialization is the process of storing objects, generally containing the data
of the simulation, to a file, such that they can be recreated by the 
application running again at some later time. The term ``serialization'' comes
from the fact that in a file data is represented as a linear stream of bytes,
and it is not necessarily trivial to determine how a complex data structure
should be put into such a form. In molecular dynamics simulations we often wish
to save the state of the system at regular intervals, perhaps every $N$ time
steps. There are two possible reasons for this: (1) We wish to defer certain
analyses until after the simulation run and (2) We wish to be able to restart a
simulation at the point where it left off, or perhaps from some intermediate
point, but changing some parameters. Apart from the atomic state (positions and
velocities, as well as the tree structure of the \code{ListOfAtoms}, 
type-names, masses etc.) it is useful to be able to save other objects in the
 system, such as the \code{Potential}, \code{Mover}, \code{NeighborLocator}, 
etc, so that if restarting at some
time in the future there will be no doubt about which parameters are associated
with which runs. Since Digital Material is intended to be run from a scripting
language such as Python, the basic parameters of a given simulation will 
typically specified in the Python script. Ideally the values of all Python
variables would be saved with the C++ objects in an automated way such that
there is never any confusion associating C++ objects to appropriate parameters.
Presently this is not the case, and applications must arrange their own manner
of coordinating the saving of basic simulation parameters and C++ objects. A 
typical method is at each time step to store the main C++ objects 
(\code{ListOfAtoms}, \code{Potential}, \code{Mover}, etc.) in one file, and 
save the Python variables using the
pickle module, in a different file but with a clearly related name (e.g. the 
C++ filename with \texttt{.pickle} appended).

\subsubsection{Serializing C++ objects in Digital Material}

We will now focus on how we serialize and de-serialize (restore from a file)
C++ objects in Digital Material. In the spirit of OOP, we have separated the
interface of the Serialization from the implementation. The interface is 
defined by three abstract base classes (two of which are closely related),
\code{Serializable}, \code{DMWriter} and \code{DMReader}. The methods of these 
classes are shown in table \ref{serializableInterface}.

\begin{sidewaystable}\label{serializableInterface}
\caption{Methods for \code{Serializable}, \code{DMWriter} and \code{DMReader}.}
\begin{tabular}{|l|l|l|}
\hline
Serializable & DMWriter & \code{DMReader} \\
\hline
 Save(DMWriter *) &  Open() &  Open() \\
 Load(DMReader *) &  Close() &  Close() \\
string \&GetType() &  Put(Serializable *, string\&) & string GetType(string \&name) \\
&  PutBool(bool, string\&) & bool Fill(Serializable *, string \&name) \\
&  PutInt(int, string \&name) & int GetInt(string \&name) \\
&  PutDouble(double, string \&name) & double GetDouble(string \&name) \\
&  PutString(string \&name, string \&name) & string GetString(string \&name) \\
&  PutIntArray(int \*, vector$\langle$int$\rangle$ \&, string \&name) & int \*GetIntArray(vector$\langle$int$\rangle$ \&, string \&name) \\
&  \ldots  &
\ldots \\

\hline
\end{tabular}
\end{sidewaystable}

Thus a \code{DMWriter} provides methods for saving the primitive data types: 
integer, double, bool, etc. as well as arrays of these. In addition there is a
\code{Put()} function which takes a pointer to a \code{Serializable} object and 
saves the state of
that object (the work will actually be done by the various Put functions). The 
\code{DMReader} provides methods for reading the same primitive types from a 
given file, as well as two functions for restoring \code{Serializable} objects:
 \code{Get()} and \code{Fill()}. The difference between them is that Get takes a 
name (a string), 
creates the appropriate class object and restores its state from the file;
 \code{Fill()} takes a
 pointer to an already created, but ``empty'' \code{Serializable} object of the
appropriate type, and ``fills in'' its member data. Note that a string-name is
associated with every piece of data that is saved/put or loaded/gotten, 
including the entire object. 

For a class to be serializable, it must derive from the base class 
\code{Ser\-ial\-iz\-able}, and thus implement the three methods in table 
\ref{serializableInterface}. Two have obvious purposes: the Save and Load 
methods, upon being passed a pointer to a \code{DMWriter} or \code{DMReader} 
respectively, 
call the latter's methods in order to save or load the individual primitive
objects making up the object's state. The GetType() function always returns a 
string equal to the name of the class. The purpose of this is to allow objects
of an appropriate type to be created given a string containing the class name.
This is straightforward in Python (using the \texttt{exec} command for example)
but requires some mechanism in C++, which for which we have used the 
Design Patterns ``Abstract Factory'' and ``Factory Method'', with Abstract 
Factory itself using the ``Singleton'' pattern. The description of the process 
is a little complex to describe, but it uses surprisingly little code. When 
high level code wishes to serialize an object, it creates a \code{DMWriter}, 
and calls its \code{Put()} method, passing the \code{Serializable} object (as 
a pointer), as  well as a
name, by which the object can be retrieved (the name is necessary since one
could save more than one object of the same type to a given file, so a means of
distinguishing them is necessary. For example the name for a \code{ListOfAtoms}
 could be \texttt{``LOA\_00012''} for the 13th (counting from zero) 
\code{ListOfAtoms} to be saved to this file. The \code{DMWriter} gets the type
 name for the \code{Serializable}
(e.g. \texttt{``DynamicListOfAtoms''}) and saves this along with the name that
 was passed. Next it calls the Save() method of the \code{Serializable}, 
passing itself. 
The Save method uses methods of the \code{DMWriter} to save all of its data.

The magic comes when we come to recreate an object from a file. Having created
an appropriate \code{DMReader}, passing the appropriate file name, we call 
\code{Get()},
passing the name that was used to save the object originally (e.g.
 \texttt{``LOA\_00012''}). The \code{DMReader} looks at the file and finds a 
string
containing the class name. The thing is now to create an object of that
type. For this two classes are used. \code{SerialFactory} is a singleton class 
(meaning only one instance of it ever exists at a time) which can take a string
containing a class name and return a pointer to a newly created object of that
type. It can do this because for each \code{Serializable} class there exists a
corresponding class \code{SerialBuilder}; the correspondence being through a 
template
argument. A global instance of the \code{SerialBuilder} is created for each
\code{Serializable} type (by a line at the top of its implementation file). The
 \code{SerialBuilder} has one chief method, called Build(). This method
dynamically creates (using the \texttt{new} operator) a new object of the same 
type as its template argument and returns the pointer. There is one more piece
 of the mechanism: When each \code{SerialBuilder} is
created, it registers itself with the (unique, \textit{a l\`{a}} Singleton) 
instance of \code{SerialFactory}, providing both the string containing the 
appropriate class name, and a pointer to itself. Thus the \code{SerialFactory} 
has a map from strings (containing class names) to pointers to objects which 
can create the corresponding objects.

To make this work, there is one more requirement of a \code{Serializable} (in 
addition to providing the three member functions listed in table
\ref{serializableInterface}: It must have a public constructor that takes no
arguments. Otherwise the corresponding \code{SerialBuilder} would not be able 
to construct one. In general not every single member
variable is saved/loaded: only those data which cannot be recreated later. For
example, when a \code{NeighborLocator} is serialized and then de-serialized 
(restored from file) the actual neighbor-lists are not saved, because these can
 be recreated when the \code{NeighborLocator} is reconnected to a 
\code{ListOfAtoms}. This brings up a point which is a general issue in 
Serialization: what do with references 
(pointers) to other objects. In Digital Material, the main object references 
are that from a \code{ListOfAtoms} to its \code{BoundaryConditions} and 
\code{NeighborLocator} (and vice versa for the latter) as well as from an 
\code{AtomsMover} to the \code{Potential}. One
cannot save the actual pointer value because this will certainly not be the 
same when the pointed-to object is recreated. One could imagine mechanisms
whereby a map ``\code{objectNamePtrs}'', mapping strings (containing object 
names) to 
pointers (to the respective objects) would exist within the simulation, as well
 as a map ``\code{nameToNameRefs}'' of strings to strings representing object
 references. nameToNameRefs would be serialized, but not objectNamePtrs, since
 the pointer values would be meaningless later. \code{objectNamePtrs} would be
 recreated as each object was de-serialized, and once \code{nameToNameRefs} was
 
de-serialized from it the pointers could be 
reset. However it is not clear that this could be done is a truly general way 
and we have decided for now to let the ``high-level'' application reconnect 
the objects by calling \code{Set\-Neigh\-bor\-Loc\-at\-or()}, 
\code{SetBoundaryConditions()}, etc., after de-serializing the respective 
objects. 

In Digital Material, we have implemented two concrete classes (pairs
of classes) as subclasses of \code{DMWriter}/\code{DMReader}. One saves 
everything in an ASCII format. The other uses the freely available binary file
 format known as NetCDF \cite{netCDFURL}.

\subsection{Graphics/Visualization}\label{graphics}

Visualization is an important tool for the analysis of MD simulation
results. Particularly in large scale simulations one does not know \textit{a
priori} what processes are going to take place (e.g. dislocation
motion). Ideally the software would be able to automatically identify defects
and extract their properties and trajectories from the atomistic data. However
we are not able to do this yet. Human visual analysis is still crucial. This
requires visualizing the atoms in such a way that a person can identify
features such as dislocations, cracks, and other defects. This generally comes
down to choosing an appropriate way to color the atoms (where ``color'' can
include transparent, i.e., leaving them out). This is turn involves
constructing functions whose values near defects are clearly distinct from 
those in defect-free regions. The simplest such
functions are the atomic energy and the (mis-)coordination number (number of 
neighbors within a cutoff distance). These two differ in their conceptual
basis, one being purely geometrical in nature, while the other depends upon the
interatomic potential. Other potential-based functions include various
components of the local (atomic) stress tensor, or in a dynamical simulation
the force magnitude (which is zero for all atoms in relaxed state of course).
 A recently introduced geometrical measure of mis-coordination is the 
``centro-symmetric deviation'' \cite{Kelchner/Plimpton/Hamilton:1998} which can
be 
 applied to materials in whose ground state lattice the neighbors of each atom
 occur in oppositely positioned pairs. The quantity computed is the sum over
 such pairs of neighbors of the square of the sum of the deviations of their
 positions from their ideal lattice positions with respect to the given
 atom. This quantity is zero for a homogeneous deformation.

In keeping with our general intent not to re-invent the wheel---namely, not to
re-implement features which have already been well implemented by others, but
rather to make use of existing freely available packages for standard tasks
(linear algebra, binary file storage) we have not developed our own
visualization package, but sought to make it easy to use existing packages. We
 have not made an extensive investigation of all the different packages
(OpenDX, VMD, etc. ) that are in use for MD visualization, but we have learned 
some things about how to incorporate visualization into the simulation of
materials.

There are two ``modes'' of visualization that may be employed in MD
simulations: \textit{real-time visualization} and \textit{post-processing
visualization}. Real-time visualization is only feasible when the system is 
small enough that the state of the system changes noticeably over the period
during which the simulator cares to observe it. However it has some
important uses: (1) demonstration applications of the software (2)
educational applications of the software and (3) debugging of scientific
applications, where if it is known that ``something bad happens'' within a 
short time of starting a simulation, visualization of the atoms can often give
 an immediate understanding of the problem (e.g., the time step was too big and
atoms ended up overlapping too much and thus the system exploded).

We have used the following three visualization tools, the first two of which 
have been implemented as Observers of the \code{ListOfAtoms}. 

\begin{description}
\item {\code{PlotAtoms}} A simple two-dimensional program written as part of 
the LASSPTools package\cite{LASSPToolsURL}. Its strength is its smooth 
presentation of real-time updates of the atomic state.
\item{\code{RasMol} \cite{RasMolURL}} A powerful program for visualizing 
molecules\---it has features for highlighting parts of proteins etc.---which is
 useful too for materials MD simulations. Its strength is its 3D rendering, 
and its facility for the user to interactively translate, zoom and rotate 
the ``molecule''. It is not particularly suitable for real-time visualization,
 and when used for such, presents a flickering image.
\item{\code{Chime} \cite{RasMolURL}} An enhancement to Rasmol, designed to run
 as a  plug-in for the Netscape web browser. Our experience with it is 
fairly limited. An advantage over bare Rasmol is
that it supports animations made from separate configuration files concatenated
into a single multiple-frame file. However it is not clear that it can handle
real time updates as well as PlotAtoms.
\end{description}

For real time visualization, say in the case of a demonstration application
with a few hundred atoms, we can attach the appropriate observer
(\code{Plot\-Atoms\-Ob\-ser\-ver}, \code{RasMolObserver}) in the Python script,
 and the
 graphics display will update every major time step (assuming \code{Notify()}
 is being called on the \code{ListOfAtoms} every major time step). When real 
time visualization is not 
practical, we can make use of an Observer which saves the state of the
\code{ListOfAtoms} (and the \code{NeighborLocator}, \code{Potential}, \ldots),
 every major time
step, or at whatever interval is considered appropriate (see Serialization,
subsection \ref{serialization} above). Using a separate
Python script we can read these snapshots from disk, attach the appropriate
Observer and display the snapshots in sequence, creating an effectively
real-time animation of the trajectory. From the same script we can create
configuration files in formats appropriate to other visualization tools if
desired. We also also perform elementary transformations of the positions such
as rotations and translations, or take subsets of the configuration (this
is necessary for PlotAtoms, but not for \code{RasMol}).

As discussed above it is crucial to be able to ``color'' the atoms in a useful
way. The process of coloring is abstracted as the \code{ColorMethod} base 
class, whose subclasses implement the coloring methods described above: 
\code{En\-er\-gy\-Co\-lor\-Meth\-od},
\code{CoordinationColorMethod}, etc. The chief requirement for subclasses is to
overload the function call operator to take a \code{ListOfAtoms} and an integer
 (an atom number) and return a double, representing a color value. Subclasses
 are also allowed to have an \code{Update()} function, taking a list of atoms, 
which is
intended to be used for calculating the colors of all atoms at once rather than
one at a time as requested by the graphics observer class. This can be 
important for efficiency in real time visualization. At this time we actually
have \code{PlotAtomsObserver} implemented both in C++ (with Python wrappers 
provided by SWIG) and in Python, which is somewhat
redundant. The pure Python implementation allows pure Python
 \code{ColorMethod}s to be
defined which is convenient, except that it may often be the case that
efficiency requires a C++ implementation. An alternative method of coloring is
according to which branch or leaf a given atom is on, which is useful when it 
is desired to indicate boundary atoms, for example. This is the default
coloring method of PlotAtomsObserver when no \code{ColorMethod} is specified
 (the user can choose which leaves have which colors, including the value -1 
for ``do not display'').

\section{\label{summaryMD}Summary}

We have given a fairly detailed exposition of our view of how to write a modern
molecular dynamics code, paying strict attention to modern software design
principles. We hope that with the descriptions provided in this paper, a 
person could implement a code more or less similar to ours, although this would
probably not be the case for parallelization and serialization, which involve
more detail than has been described here.

We would like to point out a further benefit of using Python. One of the 
``magic''
things about Python is that a function call only needs the function name to be
correct in order to work---there is no type checking. This means that if a
another MD code was written with quite different low level details, but with 
the same high level interface as the Digital Material, existing Python scripts
could be used with the other code. Python scripts which implement applications
at a high level could be shared between researchers using different core MD 
code. This highlights the importance of judicious choice of method names; in
particular, if standardization among different researchers could be achieved,
sharing of scripts for specific MD-based projects could become routine.

\section*{Acknowledgments} 
This work was financed by NSF-KDI grant No. 9873214 and NSF-ITR 
grant No. ACI-0085969. We thank Lance Eastgate, Markus Rauscher and Chuin-Shan
 Chen for useful conversations.

\bibliography{methods,silicon,misc}

\input{bibnames.sty}\hyphenation{Post-Script Sprin-ger}
\begin{thebibliography}{25}
\expandafter\ifx\csname natexlab\endcsname\relax\def\natexlab#1{#1}\fi
\expandafter\ifx\csname bibnamefont\endcsname\relax
  \def\bibnamefont#1{#1}\fi
\expandafter\ifx\csname bibfnamefont\endcsname\relax
  \def\bibfnamefont#1{#1}\fi
\expandafter\ifx\csname citenamefont\endcsname\relax
  \def\citenamefont#1{#1}\fi
\expandafter\ifx\csname url\endcsname\relax
  \def\url#1{\texttt{#1}}\fi
\expandafter\ifx\csname urlprefix\endcsname\relax\def\urlprefix{URL }\fi
\providecommand{\bibinfo}[2]{#2}
\providecommand{\eprint}[2][]{\url{#2}}

\bibitem[{\citenamefont{Bulatov et~al.}(1999)}]{Bulatov/others:1999}
\bibinfo{editor}{\bibfnamefont{V.}~\bibnamefont{Bulatov}} \bibnamefont{et~al.},
  eds., \emph{\bibinfo{title}{Multiscale Modelling of Materials, Mat. Res. Soc.
  Symp. Proc., vol 538}} (\bibinfo{publisher}{Materials Research Society},
  \bibinfo{year}{1999}).

\bibitem[{\citenamefont{Myers et~al.}(1999)\citenamefont{Myers, Arwade,
  Iesulauro, Wawrzynek, Grigoriu, Ingraffea, Dawson, Miller, and
  Sethna}}]{Myers/others:1999}
\bibinfo{author}{\bibfnamefont{C.}~\bibnamefont{Myers}},
  \bibinfo{author}{\bibfnamefont{S.}~\bibnamefont{Arwade}},
  \bibinfo{author}{\bibfnamefont{E.}~\bibnamefont{Iesulauro}},
  \bibinfo{author}{\bibfnamefont{P.}~\bibnamefont{Wawrzynek}},
  \bibinfo{author}{\bibfnamefont{M.}~\bibnamefont{Grigoriu}},
  \bibinfo{author}{\bibfnamefont{A.}~\bibnamefont{Ingraffea}},
  \bibinfo{author}{\bibfnamefont{P.}~\bibnamefont{Dawson}},
  \bibinfo{author}{\bibfnamefont{M.}~\bibnamefont{Miller}}, \bibnamefont{and}
  \bibinfo{author}{\bibfnamefont{J.}~\bibnamefont{Sethna}}, in
  \emph{\bibinfo{booktitle}{Mat. Res. Soc. Symp. Proc., Vol. 538}}
  (\bibinfo{year}{1999}), p. \bibinfo{pages}{509}.

\bibitem[{\citenamefont{Myers et~al.}(2001)\citenamefont{Myers, Cretegny,
  Bailey, Chen, Dolgert, Eastgate, Iesulauro, Ingraffea, Rauscher, and
  Sethna}}]{Myers/others:2001}
\bibinfo{author}{\bibfnamefont{C.}~\bibnamefont{Myers}},
  \bibinfo{author}{\bibfnamefont{T.}~\bibnamefont{Cretegny}},
  \bibinfo{author}{\bibfnamefont{N.}~\bibnamefont{Bailey}},
  \bibinfo{author}{\bibfnamefont{C.}~\bibnamefont{Chen}},
  \bibinfo{author}{\bibfnamefont{A.}~\bibnamefont{Dolgert}},
  \bibinfo{author}{\bibfnamefont{L.}~\bibnamefont{Eastgate}},
  \bibinfo{author}{\bibfnamefont{E.}~\bibnamefont{Iesulauro}},
  \bibinfo{author}{\bibfnamefont{A.}~\bibnamefont{Ingraffea}},
  \bibinfo{author}{\bibfnamefont{M.}~\bibnamefont{Rauscher}}, \bibnamefont{and}
  \bibinfo{author}{\bibfnamefont{J.}~\bibnamefont{Sethna}}
  (\bibinfo{year}{2001}).

\bibitem[{DMC()}]{DMCode}
DMC, \bibinfo{note}{{Digital Material Molecular Dynamics source code is
  available for download at {\tt http://www.lassp.cornell.edu/sethna/DM}},}.

\bibitem[{\citenamefont{Gamma et~al.}(1995)\citenamefont{Gamma, Helm, Johnson,
  and Vlissides}}]{Gamma/others:1995}
\bibinfo{author}{\bibfnamefont{E.}~\bibnamefont{Gamma}},
  \bibinfo{author}{\bibfnamefont{R.}~\bibnamefont{Helm}},
  \bibinfo{author}{\bibfnamefont{R.}~\bibnamefont{Johnson}}, \bibnamefont{and}
  \bibinfo{author}{\bibfnamefont{J.}~\bibnamefont{Vlissides}},
  \emph{\bibinfo{title}{Design Patterns: Elements of Reusable Object-Oriented
  Software}} (\bibinfo{publisher}{Addison-Wesley}, \bibinfo{year}{1995}).

\bibitem[{\citenamefont{Shenoy et~al.}(1999)\citenamefont{Shenoy, Miller,
  Tadmor, Rodney, Phillips, and Ortiz}}]{Shenoy/others:1999}
\bibinfo{author}{\bibfnamefont{V.~B.} \bibnamefont{Shenoy}},
  \bibinfo{author}{\bibfnamefont{R.}~\bibnamefont{Miller}},
  \bibinfo{author}{\bibfnamefont{E.~B.} \bibnamefont{Tadmor}},
  \bibinfo{author}{\bibfnamefont{D.}~\bibnamefont{Rodney}},
  \bibinfo{author}{\bibfnamefont{R.}~\bibnamefont{Phillips}}, \bibnamefont{and}
  \bibinfo{author}{\bibfnamefont{M.}~\bibnamefont{Ortiz}}, \bibinfo{journal}{J.
  Mech. Phys. Solids.} \textbf{\bibinfo{volume}{47}}, \bibinfo{pages}{611}
  (\bibinfo{year}{1999}).

\bibitem[{\citenamefont{Beazley and Lomdahl}(1994)}]{SPaSM}
\bibinfo{author}{\bibfnamefont{D.~M.} \bibnamefont{Beazley}} \bibnamefont{and}
  \bibinfo{author}{\bibfnamefont{P.}~\bibnamefont{Lomdahl}},
  \bibinfo{journal}{Parallel Computing} \textbf{\bibinfo{volume}{20}},
  \bibinfo{pages}{173} (\bibinfo{year}{1994}),
  \urlprefix\url{http://bifrost.lanl.gov/MD/MD.html}.

\bibitem[{\citenamefont{Beazley and Lomdahl}(1997)}]{Beazley/Lomdahl:1997}
\bibinfo{author}{\bibfnamefont{D.~M.} \bibnamefont{Beazley}} \bibnamefont{and}
  \bibinfo{author}{\bibfnamefont{P.~S.} \bibnamefont{Lomdahl}},
  \bibinfo{journal}{Computers in Physics} \textbf{\bibinfo{volume}{3}},
  \bibinfo{pages}{230} (\bibinfo{year}{1997}).

\bibitem[{\citenamefont{Hinsen}(2000)}]{MMTK}
\bibinfo{author}{\bibfnamefont{K.}~\bibnamefont{Hinsen}},
  \bibinfo{journal}{Journal of Computational Chemistry}
  \textbf{\bibinfo{volume}{21}}, \bibinfo{pages}{79} (\bibinfo{year}{2000}),
  \urlprefix\url{http://starship.python.net/crew/hinsen/MMTK}.

\bibitem[{\citenamefont{Bahn and Jacobsen}(2002)}]{Bahn/Jacobsen:2002}
\bibinfo{author}{\bibfnamefont{S.~R.} \bibnamefont{Bahn}} \bibnamefont{and}
  \bibinfo{author}{\bibfnamefont{K.~W.} \bibnamefont{Jacobsen}},
  \bibinfo{journal}{Computing in Science and Engineering}
  \textbf{\bibinfo{volume}{4}}, \bibinfo{pages}{56} (\bibinfo{year}{2002}).

\bibitem[{\citenamefont{Allen and Tildesley}(1987)}]{Allen/Tildesley:1987}
\bibinfo{author}{\bibfnamefont{M.~P.} \bibnamefont{Allen}} \bibnamefont{and}
  \bibinfo{author}{\bibfnamefont{D.~J.} \bibnamefont{Tildesley}},
  \emph{\bibinfo{title}{Computer Simulation of Liquids}}
  (\bibinfo{publisher}{Oxford University Press}, \bibinfo{year}{1987}).

\bibitem[{\citenamefont{Veldhuizen}(1998)}]{BLITZ++:1997}
\bibinfo{author}{\bibfnamefont{T.~L.} \bibnamefont{Veldhuizen}}, in
  \emph{\bibinfo{booktitle}{Proceedings of the 2nd International Scientific
  Computing in Object Oriented Parallel Environments(ISCOPE'98)}}
  (\bibinfo{year}{1998}).

\bibitem[{\citenamefont{Holian et~al.}(1991)\citenamefont{Holian, Voter,
  Wagner, Ravelo, Chen, Hoover, Hoover, Hammerberg, and Dontje}}]{Holian:1991}
\bibinfo{author}{\bibfnamefont{B.~L.} \bibnamefont{Holian}},
  \bibinfo{author}{\bibfnamefont{A.~F.} \bibnamefont{Voter}},
  \bibinfo{author}{\bibfnamefont{N.~J.} \bibnamefont{Wagner}},
  \bibinfo{author}{\bibfnamefont{R.~J.} \bibnamefont{Ravelo}},
  \bibinfo{author}{\bibfnamefont{S.~P.} \bibnamefont{Chen}},
  \bibinfo{author}{\bibfnamefont{W.~G.} \bibnamefont{Hoover}},
  \bibinfo{author}{\bibfnamefont{C.~G.} \bibnamefont{Hoover}},
  \bibinfo{author}{\bibfnamefont{J.~E.} \bibnamefont{Hammerberg}},
  \bibnamefont{and} \bibinfo{author}{\bibfnamefont{T.~D.}
  \bibnamefont{Dontje}}, \bibinfo{journal}{Physical Review A}
  \textbf{\bibinfo{volume}{43}}, \bibinfo{pages}{2655} (\bibinfo{year}{1991}).

\bibitem[{\citenamefont{Stillinger and Weber}(1985)}]{Stillinger/Weber:1985}
\bibinfo{author}{\bibfnamefont{F.~H.} \bibnamefont{Stillinger}}
  \bibnamefont{and} \bibinfo{author}{\bibfnamefont{T.~A.} \bibnamefont{Weber}},
  \bibinfo{journal}{Phys.\ Rev. B} \textbf{\bibinfo{volume}{31}},
  \bibinfo{pages}{5262} (\bibinfo{year}{1985}).

\bibitem[{\citenamefont{Bazant et~al.}(1997)\citenamefont{Bazant, Kaxiras, and
  Justo}}]{Bazant/others:1997}
\bibinfo{author}{\bibfnamefont{M.~Z.} \bibnamefont{Bazant}},
  \bibinfo{author}{\bibfnamefont{E.}~\bibnamefont{Kaxiras}}, \bibnamefont{and}
  \bibinfo{author}{\bibfnamefont{J.~F.} \bibnamefont{Justo}},
  \bibinfo{journal}{Phys.\ Rev. B} \textbf{\bibinfo{volume}{56}},
  \bibinfo{pages}{8542} (\bibinfo{year}{1997}).

\bibitem[{\citenamefont{Justo et~al.}(1998)\citenamefont{Justo, Bazant,
  Kaxiras, Bulatov, and Yip}}]{Justo/others:1998}
\bibinfo{author}{\bibfnamefont{J.~F.} \bibnamefont{Justo}},
  \bibinfo{author}{\bibfnamefont{M.~Z.} \bibnamefont{Bazant}},
  \bibinfo{author}{\bibfnamefont{E.}~\bibnamefont{Kaxiras}},
  \bibinfo{author}{\bibfnamefont{V.~V.} \bibnamefont{Bulatov}},
  \bibnamefont{and} \bibinfo{author}{\bibfnamefont{S.}~\bibnamefont{Yip}},
  \bibinfo{journal}{Phys.\ Rev. B} \textbf{\bibinfo{volume}{58}},
  \bibinfo{pages}{2539} (\bibinfo{year}{1998}).

\bibitem[{\citenamefont{Jacobsen et~al.}(1996)\citenamefont{Jacobsen, Stoltze,
  and N{\o}rskov}}]{Jacobsen/Stoltze/Norskov:1996}
\bibinfo{author}{\bibfnamefont{K.~W.} \bibnamefont{Jacobsen}},
  \bibinfo{author}{\bibfnamefont{P.}~\bibnamefont{Stoltze}}, \bibnamefont{and}
  \bibinfo{author}{\bibfnamefont{J.~K.} \bibnamefont{N{\o}rskov}},
  \bibinfo{journal}{Surf. Sci.} \textbf{\bibinfo{volume}{366}},
  \bibinfo{pages}{394} (\bibinfo{year}{1996}).

\bibitem[{\citenamefont{Baskes}(1992)}]{Baskes:1992}
\bibinfo{author}{\bibfnamefont{M.~I.} \bibnamefont{Baskes}},
  \bibinfo{journal}{Phys.\ Rev. B} \textbf{\bibinfo{volume}{46}},
  \bibinfo{pages}{2727} (\bibinfo{year}{1992}).

\bibitem[{\citenamefont{Holian et~al.}(1990)\citenamefont{Holian, DeGroot,
  Hoover, and Hoover}}]{Holian/others:1990}
\bibinfo{author}{\bibfnamefont{B.~L.} \bibnamefont{Holian}},
  \bibinfo{author}{\bibfnamefont{A.~J.} \bibnamefont{DeGroot}},
  \bibinfo{author}{\bibfnamefont{W.~G.} \bibnamefont{Hoover}},
  \bibnamefont{and} \bibinfo{author}{\bibfnamefont{C.}~\bibnamefont{Hoover}},
  \bibinfo{journal}{Physical Review A} \textbf{\bibinfo{volume}{41}},
  \bibinfo{pages}{4552} (\bibinfo{year}{1990}).

\bibitem[{CampOS()}]{campos}
CampOS, \emph{\bibinfo{title}{{CAMP} open software project homepage}},
  \bibinfo{note}{{\tt http://www.fysik.dtu.dk/CAMPOS}}.

\bibitem[{mpiPython()}]{mpipython}
mpiPython, \bibinfo{note}{this has been included with Konrad Hinsen's set of
  Python modules called Scientific Python, available from
  http://starship.python.net/~hinsen/ScientificPython/}.

\bibitem[{NetCDF()}]{netCDFURL}
NetCDF, \emph{\bibinfo{title}{{NetCDF} home page}},
  \urlprefix\url{http://www.unidata.ucar.edu/packages/netcdf}.

\bibitem[{\citenamefont{Kelchner et~al.}(1998)\citenamefont{Kelchner, Plimpton,
  and Hamilton}}]{Kelchner/Plimpton/Hamilton:1998}
\bibinfo{author}{\bibfnamefont{C.}~\bibnamefont{Kelchner}},
  \bibinfo{author}{\bibfnamefont{S.}~\bibnamefont{Plimpton}}, \bibnamefont{and}
  \bibinfo{author}{\bibfnamefont{J.}~\bibnamefont{Hamilton}},
  \bibinfo{journal}{Phys. Rev. B} \textbf{\bibinfo{volume}{58}},
  \bibinfo{pages}{11085} (\bibinfo{year}{1998}).

\bibitem[{\citenamefont{Sethna et~al.}(1990)}]{LASSPToolsURL}
\bibinfo{author}{\bibfnamefont{J.~P.} \bibnamefont{Sethna}}
  \bibnamefont{et~al.}, \emph{\bibinfo{title}{{LASSPTools: Graphical and
  Numerical Extensions to Unix}}} (\bibinfo{year}{1990}),
  \urlprefix\url{http://www.lassp.cornell.edu/LASSPTools/LASSPTools.html}.

\bibitem[{RasMol()}]{RasMolURL}
RasMol, \emph{\bibinfo{title}{{RasMol} home page}},
  \urlprefix\url{http://www.umass.edu/microbio/rasmol}.

\end{thebibliography}

\end{document}